\documentclass[showkeys,prl,amsmath,amssymb,floatfix,twocolumn
]{revtex4}
\usepackage{graphicx}
\usepackage{dcolumn}
\usepackage{bm}
\usepackage{natbib}
\usepackage{epstopdf}
\usepackage{natbib}
\usepackage{float}
\usepackage{amsmath} 
\usepackage[abs]{overpic}
\usepackage{subfigure}
\usepackage{color} 
\usepackage{tabularx}
\newcommand{\upperRomannumeral}[1]{\uppercase\expandafter{\romannumeral#1}}

\begin{document}
\title{Wing fold angle determines the terminal descent velocity of double winged autorotating seeds, fruits and other diaspores}

\author{Richard A. Fauli, Jean Rabault and Andreas Carlson}
 \email{acarlson@math.uio.no}
%
%
\affiliation{Department of Mathematics, Mechanics Division, University of Oslo, 0316 Oslo, Norway.}
\date{\today}

\begin{abstract}
Wind dispersal of seeds is an essential mechanism for plants to proliferate and to invade new territories. In this paper we present a methodology that combines 3D-printing, a minimal theoretical model, and experiments to determine how the curvature along the length of the wings of autorotating seeds, fruits and other diaspores provides them with an optimal wind dispersion potential, i.e., minimal terminal descent velocity. Experiments are performed on 3D-printed double winged synthetic fruits for a wide range of wing fold angles (obtained from normalized curvature along the wing length), base wing angles and wing loadings to determine how these affect the flight. Our experimental and theoretical models find an optimal wing fold angle that minimizes the descent velocity, where the curved wings must be sufficiently long to have horizontal segments, but also sufficiently short to ensure that their tip segments are primarily aligned along the horizontal direction. The curved shape of the wings of double winged autorotating diaspores may be an important parameter that improves the fitness of these plants in an ecological strategy. 
\end{abstract}




\maketitle



\section{Introduction}
Wind dispersion of seeds is a widespread evolutionary adaptation found in plants, which allows them to multiply in numbers and to colonize new geographical areas \cite{Nathan786,bookRidley,Howe1982,Nathan2008, cain2000,green1989}. Seeds, fruits and other diaspores (dispersal units) are equipped with appendages that help generate a lift force to counteract gravity as they are passively transported with the wind. Seeds with a low terminal descent velocity increase their flight time and the opportunity to be transported horizontally  by the wind before reaching the ground \cite{Tackenberg2003}. Many plant species are today unfortunately under severe stress and on the verge of becoming extinct due to climate change, timber extraction and agricultural development \cite{bookghazoul}. The terminal velocity of the seed is a necessary prerequisite for accurate predictions from dispersion models \cite{Nathan786,Tamme2014}, which can help predict their wind dispersion and influence policy-makers in their conservation and reforestation plans \cite{bookghazoul}.

Since wind dispersal of seeds occupies a critical position in plant ecology, their flight organs have been carefully described along with their flight pattern \cite{Augspurger1986, norberg1973autorotation, azuma1989flight}. These flight organs are often leaf-like structures that function as wings, allowing the seed or diaspore to autorotate \cite{norberg1973autorotation}, tumble or glide \cite{Augspurger1986} as it is pulled to Earth by gravity. Other flight solutions are composed of thin-hairy structures such as the pappus on the dandelion \cite{Cummins2018}, which effectively serves as a parachute. Common to all of these are the fact that their dispersion mechanisms rely on mechanical principles once they are released from the mother plant, a trait shared across plant species \cite{Elbaum884, Marmottant20131465,Skotheim1308,Noblin1322,Armon1726}. 

Single bladed autorotating seeds are often associated with maple trees and conifers, where the seed is attached to a single straight wing \cite{azuma1989flight}. The delicate balance between the weight of the seed and the shape of the wing allows it to autorotate \cite{varshey2012}, leading to the production of an unexpectedly high lift \cite{azuma1989flight}. Measurements of the air flow produced around autorotating {\em samaras} identify a leading edge vortex that is primarily responsible for the production of a positive lift force \cite{lentink2009leading,lee2014mechanism}. The seemingly simple configuration of having a single wing that generates a stable rotary descent has been widely studied \cite{smith1971,maha1999, pesavento2004}, where recent work has shown that also the wing elasticity can influence the flight pattern \cite{tam2010} and may enhance lift \cite{Wang2013,tam2015}. Fossils from voltzian conifers dating back to the late early to middle Permian (ca. 270 Ma) are found to be double-winged \cite{Stevenson2015}. {This wing geometry is today vastly outnumbered by the autorotating single-winged morphology in the same plant family, which suggests that in the context of an ecological strategy the flight performance of single- winged seeds improves the fitness of their producers \cite{Stevenson2015,Contreras2015}. Pollen from the genus {\em Pinus} \cite{cain1940} have also been suggested to have evolved into shapes that improve their aerodynamic performance.}

Autogyrating motion is also widely observed in multi-winged diaspores and seeds. These are commonly known as whirling fruits or helicopter fruits, which can be found in plant families such as Dipterocarpaceae \cite{smith2015predicting,suzuki1996sepal, Matlack}, Hernandiaceae, Rubiaceae \cite{VANSTADEN1990542} and Polygonaceae, occurring in Asia, Africa and the Americas. These fruits are equipped with a leaf like structure (persistent and enlarged sepals), which acts as wings in their rotary descent, illustrated  in their Greek name, i.e., di = {\em two}, pteron = {\em wing} and karpos = {\em fruit}. 
 Compared to the single bladed maple fruits, these have a more complex wing shape which curves upwards and outwards \cite{nla.cat-vn610737}. Only a limited sub-set of tropical whirling fruits are described in terms of their terminal descent velocity as illustrated by the data from 34 neotropical trees \cite{Augspurger1986}, $53$ recodings by \cite{Tamme2014} and recently extended by 16 entries of Paleotropic trees \cite{smith2015predicting}, which clearly limits predictions of dispersal distance. These flight recordings \cite{Augspurger1986,smith2015predicting} suggest that the descent velocity is proportional to the square-root of the wing-loading, i.e., its mass divided by the disk defined by the projected wing area during rotation \cite{green1989}. Recordings of the rotational frequency of this class of multi-winged fruits are elusive and essential to get a complete understanding of their aerodynamics. There are no studies, to the best of our knowledge, that characterize how the wing shapes of double-winged whirling fruits influence the terminal descent velocity and their rotational frequency. To understand the relationship between the wing geometry and the terminal descent velocity we deploy a methodology that combines 3D printing of synthetic fruits, experiments and a minimal theoretical description of their flight based on the blade element theory. 
\begin{figure}
 \begin{center}
  \includegraphics[width=.45\textwidth]{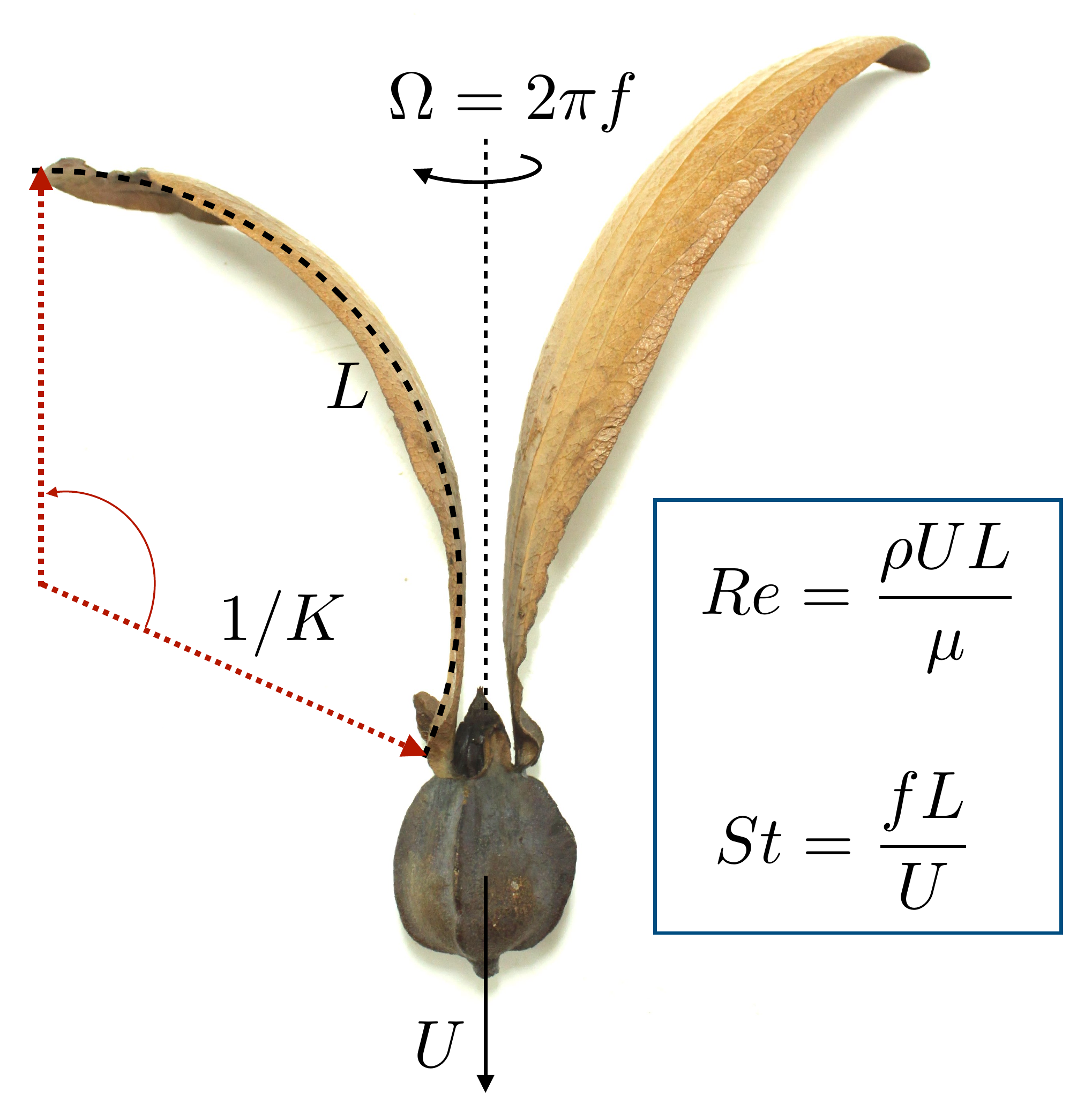}
  \caption{\label{sketch} We parameterize the shape of the fruit and its sepals, here illustrated by a dried fruit and wings from the dipterocarp family where we extract the sepal length ($L$) and notice that the sepal has a curvature $K$ along the length of the wing here illustrated by fitting an osculating circle along the wing giving its radius of curvature $1/K$. The sepal's fold angle is defined as $KL$. The fluid has a density $\rho$ and a viscosity $\mu$. When the fruit is in free fall, it descends with a terminal velocity $U$ and autogyrates with an angular frequency $\Omega=2\pi f$, where $f$ is the rotational frequency. Dimensional analysis gives us in addition to the fold angle $KL$, two non-dimensional numbers that describe the flight; the Reynolds number ($Re$) is giving the ratio of inertia and the viscous force $Re=\frac{\rho U L}{\mu}$ and the Strouhal number ($St$) is giving the ratio of the rotational speed and the translational speed $St=\frac{fL}{U}$. Image courtesy of James Smith.}
 \end{center}
 \end{figure}

\section{Methodology}
\subsection{Scaling analysis}
One example of a whirling fruit from the dipterocarp family is shown in Fig. 1, which will autorotate as it descends. The angular frequency is $\Omega=2\pi f$, where $f$ is the rotational frequency and $U$ is the terminal descent velocity. The sepals in Fig. 1 have a length $L$ and a curvature $K$ and the fruit is pulled to Earth by gravity with an acceleration $g$. Through dimensional analysis we define three non-dimensional numbers; the Reynolds number ($Re$) is giving the ratio of inertia and the viscous force $Re=\frac{\rho U L}{\mu}$, the Strouhal number ($St$) is giving the ratio of the rotational speed and the translational speed $St=\frac{fL}{U}$ and the wings fold angle $KL$ in radians. The fluid has a density $\rho$ and a viscosity $\mu$. The experiments and theoretical analysis are based on a flow with $Re\gg1$ and for steady descent, i.e., $U=constant$, consistent with previous experiments on autorotating multi-winged fruits collected in the wild with $Re \in 10^3-10^4$ \cite{norberg1973autorotation,Augspurger1986, smith2016predicting} (see Table \ref{table_scaling}). As  the flow is dominated by inertia, we know then that the lift force scales as $\vert F_L\vert \sim C_L \rho U^2 A_d$ and the drag force scales as $\vert F_D\vert \sim C_D \rho U^2A_d$ where $C_L$ is the lift coefficient, $C_D$ the drag coefficient and $A_d$ is the area swept by the wing. Assuming force balance between the gravitational force and the lift force leads to the scaling prediction, usually formulated as \cite{norberg1973autorotation,BURROWS1974}:
\begin{equation} 
\label{eq:scaling}
U\sim\sqrt{mg / A_d},
\end{equation} 
where we have omitted the effective lift coefficient, {which depends on the exact geometry of the fruit}, and the fluid density. {Therefore, to capture the details of this coefficient a more detailed analysis is required, either by using a phenomeological model such as the blade element model, or resolving the in flow a numerical simulation.} Experiments on wild fruits show that individual groups \cite{Augspurger1986,smith2015predicting} follow the scaling law Eqn. (\ref{eq:scaling}), but they lack a description of how $U$ is affected by the wing geometry. 

\subsection{Synthetic double-winged fruits}
\begin{figure}[]
 \begin{center}
  {\includegraphics[width=.52\textwidth]{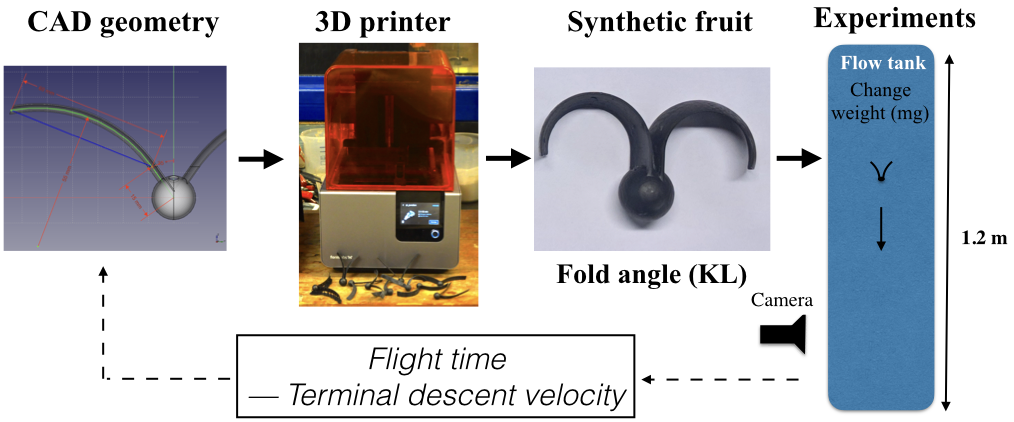}}

  \caption{\label{baseline_model} Our experimental methodology and work flow is composed of digitalization of the synthetic fruits by a 3D Computer Aided Design model (FreeCAD v0.16), which is then 3D-printed and used for flight experiments in a water tank. After varying the weight $mg$ of each geometry by depositing a known weight of lead to its hollow fruit, we measure the terminal descent velocity $U$ and rotational frequency $f$. After data evaluation we return to the CAD design with guidelines for the wing design, i.e., most significantly in this study the fold angle $KL$.}
 \end{center}
 \end{figure}

The workflow of our experimental approach is illustrated in Fig. \ref{baseline_model}. It consists in producing rapid prototyping synthetic seeds through 3D-printing, then experimenting in a water tank to extract the terminal descent velocity ($U$) and the rotational frequency ($f$), which is evaluated and then fed-back to the design of a new wing geometry. A parametric 3D Computer Aided Design model (CAD) is developed in FreeCAD v0.16 \cite{FreeCAD}, see right part in Fig. \ref{baseline_model} and CAD file (see Supplemental Material). { The length $L$ of the curved part of the wing is kept constant and equal to $6$ cm. The fold angle $KL$, which is the geometrical parameter given as an input to the CAD model and can be modified while keeping the total wing mass constant. The wing camber is set through a radius of curvature in the plane normal to the wingspan, and can also be adjusted without influencing the wing mass. We set a value for an effective additional angle of attack of $2.5^{\circ}$ (see Eqn. (\ref{added_lift})) and the wing pitch angle $\alpha_p = 15^\circ$, inspired by the geometry of the wings of wild fruits. It is very challenging to accurately measure the pitch and camber values from wild fruits, as these are strongly affected by the desiccation process and their growth. The angle $\psi$ between the base of the wing and the vertical direction is chosen equal to $\psi=0^{\circ}$ or $\psi=35^{\circ}$, which encompasses the typical range of values observed in nature \cite{PRL}. Finally, the fruit itself is designed as a hollow sphere with two holes, allowing us to vary its weight in experiments while keeping the volume fixed.}

 The synthetic fruits are produced by using the Form 2 3D-printer from Formlabs \cite{formlabs}, relying on the stereolithography technique with a print time of about 5 hours for a batch of 3 fruits. This allows for rapid prototyping, and parametrization of fruit geometry as required for scanning a large phase space of shapes, see Fig. \ref{baseline_model}. Once the model is printed, it is cleaned in isopropanol, cured in UV light and polished to have a smooth surface before being used for experiments in the water tank.

 \subsection{Experimental design}
 We design our experiments so that both the Reynolds number ($Re$) and the Strouhal number ($St$) are in direct correspondence to fruit flight in Nature. The values for these parameters in our laboratory experiments correspond to those performed on wild fruits and are reported in Table \ref{table_scaling}.

 Experiments were performed in a cylindrical water tank of a height $1.2$~m and a diameter $25$~cm. A set of experiments were performed in a large water tank filled with water of depth $0.7$~m, $0.5$~m wide and $10$~m long, to make sure that wall effects are within the experimental error bars in the cylindrical tank. {In the experiments, we control the amount of lead added to the hollow spherical fruit of the 3D-printed model and the additional volume is filled with water. The fruit is fully immersed under the water surface before it is released.} 
 
 A camera is recording the motion of the fruit and wings from the side of the tank at a frequency of $30$Hz and a resolution of 864 $\times$ 480 pixels. Images are extracted from the video to track the fruit's lowest point and the wing tips. In our data analysis, we subtract the background to each image in the video, which makes the fruit to easily be identified and a combination of convolution filtering and thresholding is used to find the characteristic points, see Fig. \ref{tracking} and the code for more details (see Supplemental Material). Calibration resorting to a third order polynomial is used to convert the position of the characteristic points in each image into the vertical position. The fruits rotational frequency is obtained by using a Fast Fourier Transform of the position of the wing tips, where the frequency is identified as the peak in the power-spectrum. To verify our post-processing analysis, we also performed measurements looking from the top and down into the water tank, which are found to be in excellent agreement with the measurements obtained by the side view. Each experiment is repeated ten times for ${\psi} = 0^{\circ}$ and three times for ${\psi}=35^{\circ}$. All the reported data points are average values and the error bars is five times the standard deviation.

 \begin{table*}[ht!]
 \begin{center}
  \begin{tabular}{ccccccc}
  \hline
  Setup & $f$ (Hz) & $L$ (m) & $U$ (m/s) & $\nu$ (m$^2$/s) & $St$ & $Re$ \\
  \hline\hline
  laboratory   & 0.4 to 2.4 & 0.06 & 0.04 to 0.2 & $1.0 \cdot 10^{-6}$ & 0.35 to 0.82 & 2400 to 14000 \\
  \hline
  {\em Shorea argentifolia} Symington (Dipterocarpaceae) & N/A & $\approx 0.035$ & 1.2 & $1.5 \cdot 10^{-5}$ & 0.40 ($^*$) & 2800 \\
  \hline
 {\em Shorea johorensis} Foxw. (Dipterocarpaceae) & N/A & $\approx 0.05 $ & 1.7 & $1.5 \cdot 10^{-5}$ & 0.40 ($^*$) & 5700 \\
  \hline
   {\em Shorea mecistopterix} Ridl. (Dipterocarpaceae) & N/A & $\approx 0.08$ & 2.4 & $1.5 \cdot 10^{-5}$ & 0.40 ($^*$) & 12700 \\
  \hline
  \end{tabular}
  \caption{Description of the parameters representative for the laboratory experiments and comparison with previously performed experiments on wild fruits from \cite{smith2016predicting}, where the rotation frequency $f$ is not reported. The typical length of the sepals of wild fruits is estimated as $L = \sqrt{A_w/2}$, where $A_w$ is the wing area reported in \cite{smith2016predicting}. The value for $St$ is estimated from \cite{azuma1989flight} in the case of {\em Buckleya lanceolata} (Sieb. \& Zucc.) Miq. (syn. {\em B. joan} (Sieb.) Makino (Santalaceae)) with a geometry similar to a {\em Dipterocarpus} fruit giving us the $St$ number. The $^*$ symbol indicates that the numerical value could not be computed from the data reported in \cite{smith2016predicting}, but obtained from a species of similar shape \cite{azuma1989flight}.}
  \label{table_scaling}
 \end{center}
 \end{table*}

\begin{figure}
 \begin{center}
  \includegraphics[width=.45\textwidth]{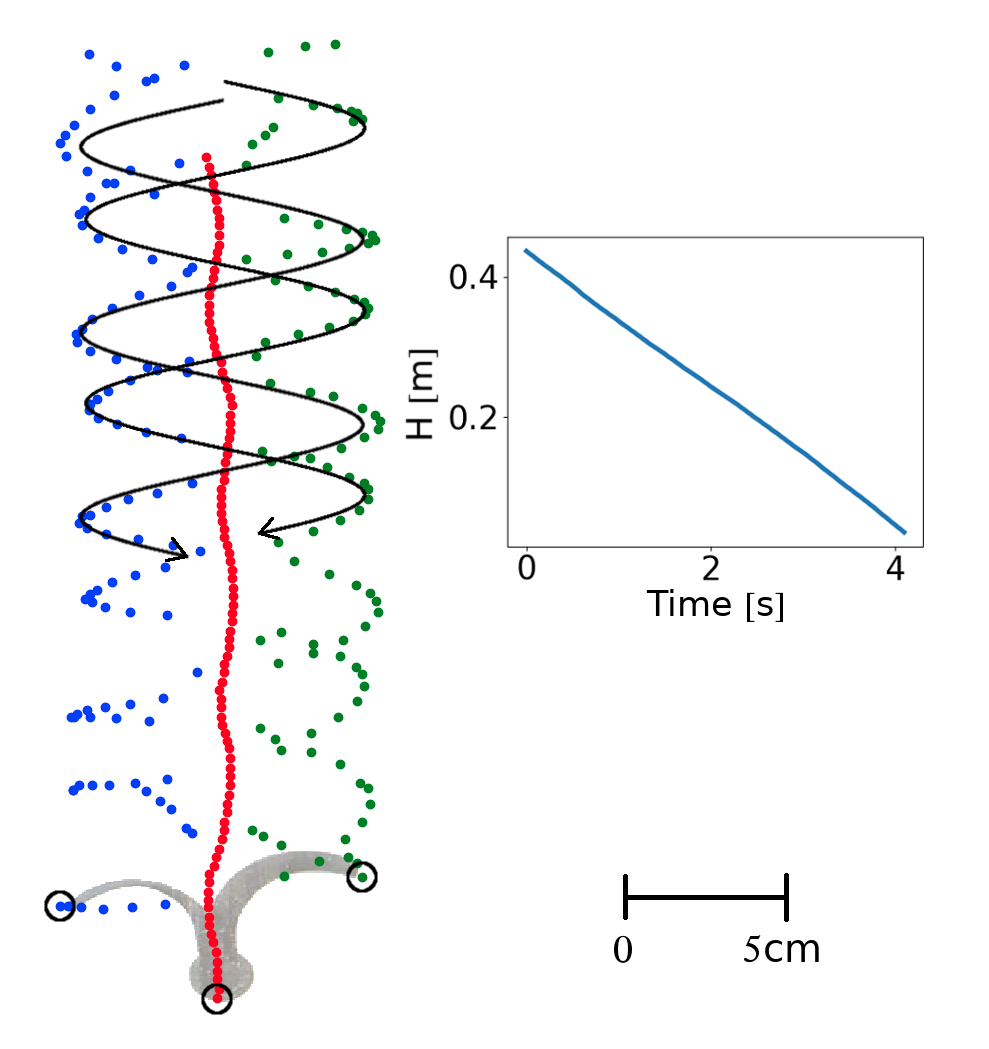}
  \caption{\label{tracking}The left part shows the raw data from our automatic tracking of the motion of a synthetic fruit, where we follow its wing tips and the vertical motion of the fruit. The lines are intended only as a guide to the eye to illustrate the motion of each individual wing. In the right part, we plot the vertical position {$H$} of the fruit in one part of the tank and show that our synthetic fruits reach a steady terminal descent velocity $U$.}
 \end{center}
 \end{figure}

\subsection{Blade element model}
The blade element model describes the lift and torque generated by rotating wings \cite{norberg1973autorotation,azuma1989flight,lee2014mechanism} by considering the wing as a succession of small elements. For each element the relative wind, the local angle of attack, the lift force and the drag force are computed.

\begin{figure*}
 \begin{center}
  \subfigure[\label{seed_decomposition}]{\includegraphics[width=.45\textwidth]{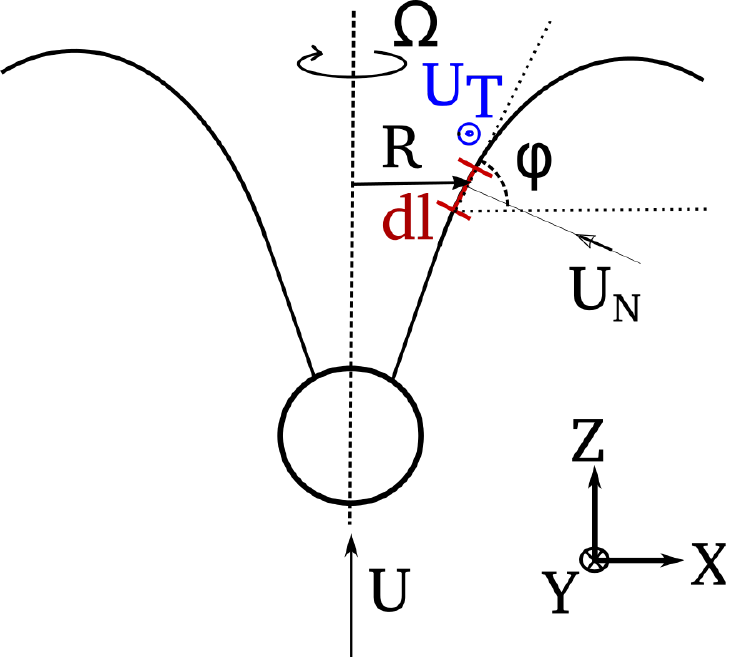}}
  \subfigure[\label{wing_element}]{\includegraphics[width=.45\textwidth]{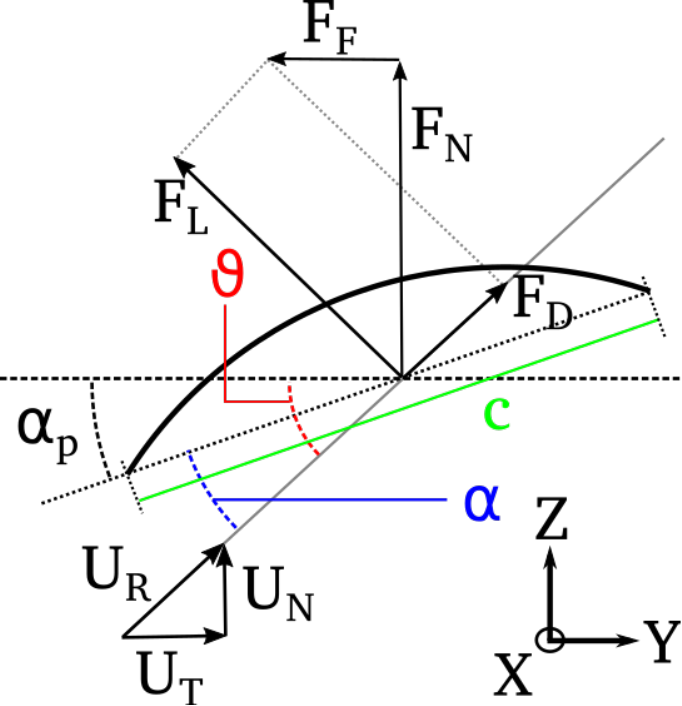}}
  \caption{(a) Sketch of the fruit's wings that are decomposed into small elements $dl$ and moving with a frame of reference at the vertical descent velocity $U$. $\Omega$ is the rotation rate, $dl$ the length of the blade element, $R$ the distance of the blade element to the axis of rotation, $\phi$ the wing inclination at the location of the blade element, $U_N = U \cos(\phi)$ the normal component of the descent velocity, $U_T = R \Omega$ the tangential wind created by the rotation. (b) View of the blade element described in Fig. \ref{seed_decomposition}, in the plane perpendicular to the wingspan. $c$ is the chord of the blade element, $\alpha_p$ its pitch angle, $\theta$ the angle of the relative wind, $\alpha = \theta - \alpha_p$ the angle of attack, $U_R$ the wind speed ''felt'' by the blade element, $L$ and $D$ the lift and drag components, respectively, and $F_F$ and {$F_N$} the resulting forward and vertical forces. The camber profile is illustrated by the curvature of the blade element.}
 \end{center}
 \end{figure*}

 The rotating fruit is parametrized (see Fig. \ref{seed_decomposition}), with $U$ the fruit's vertical descent speed, and $\Omega$ the rotation rate. For a given blade element of length $dl$, the inclination of the wing relative to the horizontal direction is $\phi$ so that the descent velocity has a projected component $U_N = U \cos(\phi)$ normal to the blade element. $R$ is the distance between the blade element and the fruit's axis of rotation, so that the tangential velocity becomes $U_T = R \Omega$. In addition, $\psi$ is the angle between the vertical direction and the wing base. $\psi = 0^\circ$, unless stated otherwise.

%

 A view of the relative wind and the aerodynamic forces acting on the blade element in the plane perpendicular to the wingspan is shown in Fig. \ref{wing_element}. $\alpha_p$ is the pitch angle of the blade element, $c$ its chord, and the camber is visible through the curvature of the wing profile. The total relative wind at the blade element has a magnitude $U_R = \sqrt{U_T^2 + U_N^2}$, and is at an angle $\theta = \arctan(U_N / U_T)$ relative to the horizontal direction where the blade element's angle of attack is $\alpha = \theta - \alpha_p$. The lift force is the component of the aerodynamic force perpendicular to the direction of the relative wind, while drag is the force component parallel to the wind direction. The magnitude of the lift force $F_L = \frac{1}{2} \rho U_R^2 A C_L(\alpha)$ and the drag force $F_D = \frac{1}{2} \rho U_R^2 A C_D(\alpha)$ on an element are proportional to the lift and drag coefficients, $C_L(\alpha)$ and $C_D(\alpha)$ where $A = c \cdot dl$ is the area of the blade element. 


 The horizontal (forward) and {normal forces, respectively $F_F$ and $F_N$}, can therefore be obtained by projecting the total aerodynamic force obtained by the sum of lift and drag:

 \begin{equation}
  \begin{aligned}
    F_F &= F_L \sin(\theta) - F_D \cos(\theta)\\
    F_N &= F_L \cos(\theta) + F_D \sin(\theta).
  \end{aligned}
\end{equation}

  As shown in Fig. \ref{wing_element}, the horizontal force can be directed forward and act as a motor for the rotation of the wings. This takes place when the angle of the relative wind is large enough so that the forward projected component of the lift force exceeds the backward component of the drag force, {i.e., close to the central axis of the fruit.} This forward resultant force generates autorotation of winged fruits, in a similar way to what is used on a helicopter in autorotation \cite{shapiro1956principles}. As a consequence, a moment that drives autorotation is produced close to the axis of rotation of the fruit, while a moment that opposes rotation is produced close to the wing tips where the tangential velocity, and therefore drag, is the dominant horizontal force \cite{norberg1973autorotation}. {These quantities are further illustrated by the results shown in Fig. \ref{example_distribution}}.

 The total vertical force and torque acting on the seed are obtained by integrating along the two wingspans and adding the effect of gravity, i.e., writing $F$ the resulting vertical force, and $M$ the moment around the rotational axis of the fruit

 \begin{equation}
 \begin{aligned}
  F &= 2 \int_{l=0}^{l=S} F_N \cos(\phi) dl -mg\\
  M &= 2 \int_{l=0}^{l=S} F_F R dl,
 \end{aligned}
 \label{Eqn:force_moment}
 \end{equation}

\noindent where $S$ is the total curvilinear length along the wingspan of one wing, $R$ is the distance between the blade element and the vertical axis of the fruit, $m$ the mass relative to the surrounding fluid, and $g$ the acceleration of gravity.
%
 
\begin{figure*}[ht!]
 \subfigure[ ]{\includegraphics[width=.4\textwidth]{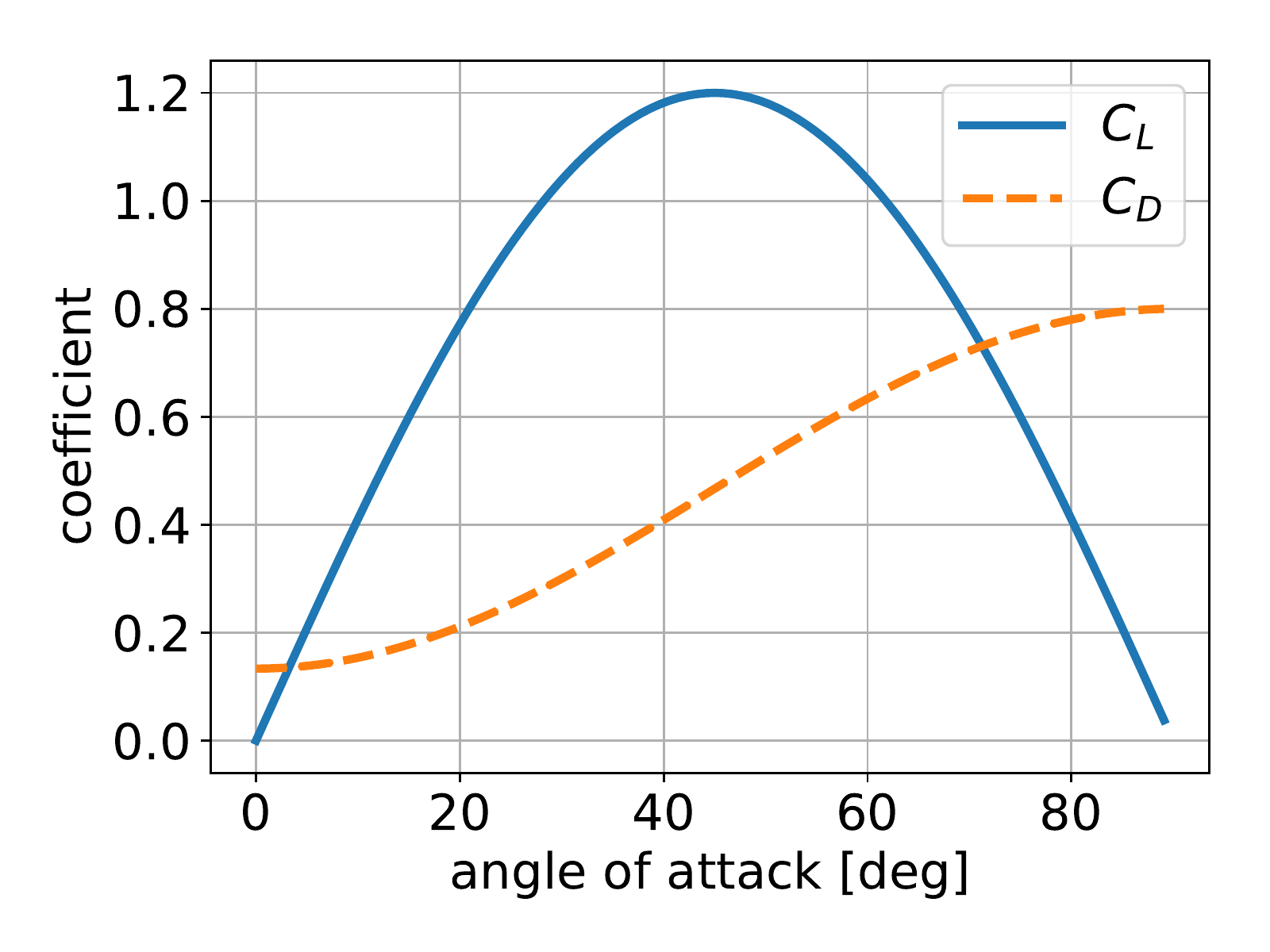}}
 \subfigure[ ]{\includegraphics[width=.4\textwidth]{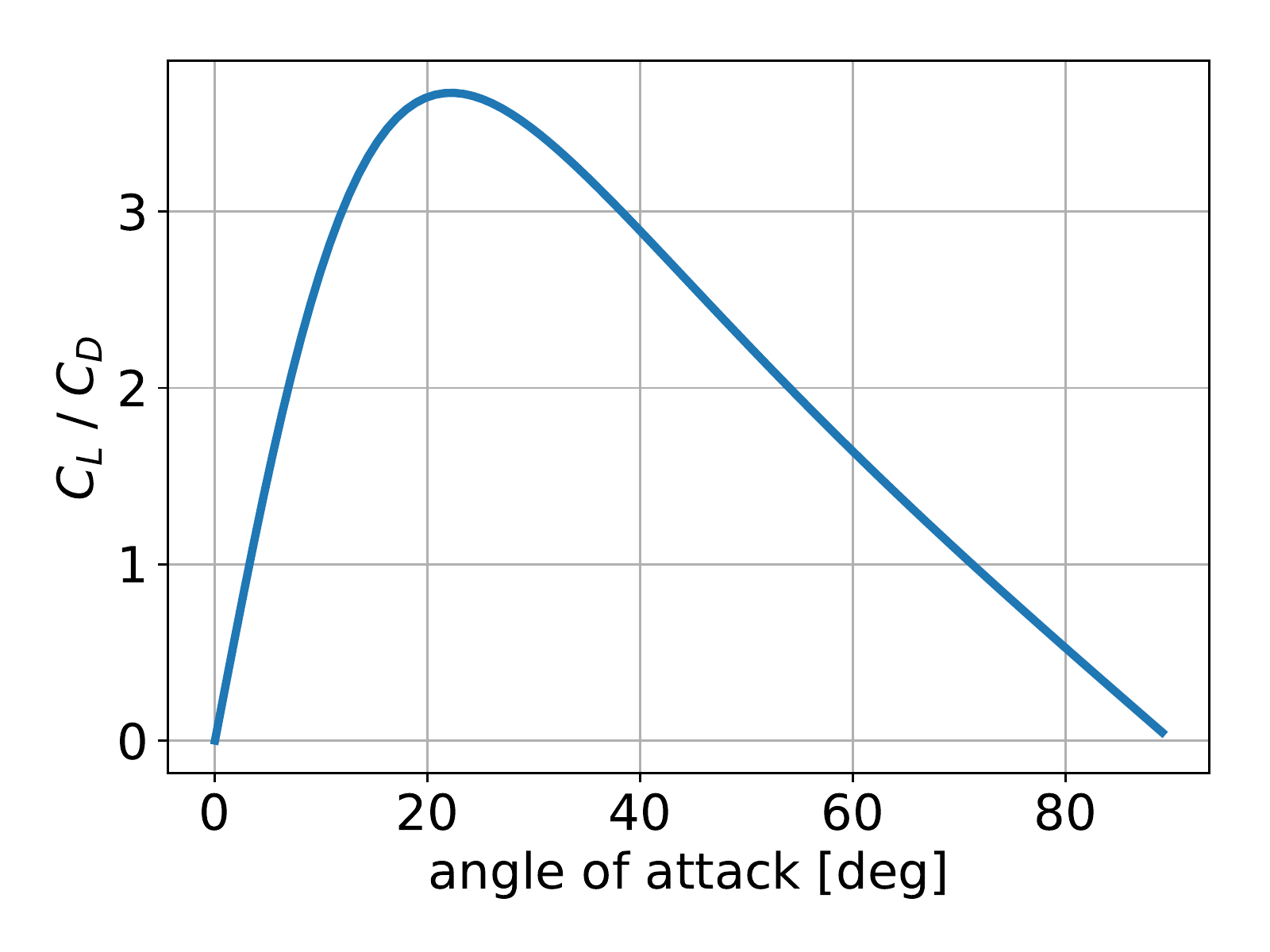}}
 \caption{\label{CL_o_CD} { (a) Parametrization used for the $C_L(\alpha)$ (solid line) and $C_D(\alpha)$ (dashed line) coefficients. A much smoother stall behavior is modeled compared with high Reynolds number wings, in agreement with studies performed for rotating wings at moderate Reynolds numbers where a strong leading edge vortex is present. (b) The parametrization results in a smooth shape for the $C_L(\alpha) / C_D(\alpha)$ curve.} }
 \end{figure*}
 
 Therefore, the predictions of the blade element model depend on two set of parameters; {most importantly, the geometry of the wing, and also the lift and drag coefficients as a function of the angle of attack. This second part is arguably the most challenging to model accurately, as it is difficult to find tabulated values for the drag and lift coefficients of rotating wing segments at low to intermediate Reynolds numbers. This may come from a variety of reasons, including the small values of the forces acting on wing segments at the corresponding scales and Reynolds numbers in either air or water, which makes experimental measurements challenging. We have used two sets of parametrizations for the lift and drag coefficients; one from lift experiments on translating wing at slightly higher Reynolds numbers, and one inspired from rotating and flapping wings at similar Reynolds numbers as compared with our study. Both parameterizations qualitatively captures the trends in our experimental measurements.}
 
 For flat plates at high Reynolds number, the lift and drag coefficients can be obtained from both theoretical considerations (inviscid fluid flow with circulation together with the Kutta-Joukowski theorem, and boundary layer friction \cite{landau2013fluid}), and experiments. These show that the lift force is a function of the angle of attack, i.e., $C_L(\alpha) = 2 \pi \sin(\alpha)$ \cite{landau2013fluid} and accurate until stall occurs, which for a flat plate usually arises at around $\alpha = 6^{\circ}$. In the case of low to intermediate Reynolds numbers, the behaviour of $C_D(\alpha)$ and $C_L(\alpha)$ needs to be modified and we follow the results of \cite{lissaman1983low}, by using a maximum value of typical value $\max(C_L(\alpha)/C_D(\alpha))=7$ before stall. {This parametrization gives results in good agreement with the experimental terminal descent velocity as a function of fold angle $KL$, and are also robust to changes in the peak value of $C_L / C_D$ (see Supplemental Material). The Strouhal number around the optimal fold angle is also in good agreement with experiments. However, when one goes further away from the optimal fold angle, we find the stall behavior is likely too aggressive which explains the deviation from the experimental values for St.}
 
 {Indeed, experiments and simulations at Reynolds numbers comparable to the ones we experience indicate the existence of a strong leading edge vortex, that changes the stall pattern compared with higher Reynolds number \cite{Wang449, Lentink2705}. Inspired by this observation we have also used another parametrization based on data from flapping wings reported by \cite{Wang449}, where no sharp stall is present and the lift coefficient increases up to a higher angle of attack than what is observed for flat plates at higher Reynolds numbers. However, compared to the results of \cite{Wang449}, we reduce the value of the drag coefficient to obtain good agreement with our experiments. We believe this stems from the flow physics, as the aspect ratio of the wings in the corresponding work is much lower than in our work and should generate more induced drag at the wing tips. In addition, the results presented here are for complete wings rather than an individual wing element, and therefore the relative values of drag obtained in \cite{Wang449} and similar works are probably larger than what is the case on individual wing elements far from the axis of rotation of our fruits. The parametrization is summarized in Fig. \ref{CL_o_CD}. 
We will only present results obtained from using the  blade element model with a parametrization according to Fig. \ref{CL_o_CD} in the main text and the additional simulations are in the Supplemental Material, where we note that a precise curve for $C_D(\alpha)$ and $C_L(\alpha)$ would require a detailed flow measurements or high resolution computational fluid dynamics simulations, beyond the scope of this article.}
 
 
 We note that the wing camber also affects the lift generation by adding an offset to the lift coefficient curve \cite{newman1977marine}, $C_{L, camber}(\alpha) = C_{L, no camber}(\alpha + \Delta \alpha)$. While this result applies primarily for higher Reynolds numbers than what we consider here, this is used as a first approximation of the behavior expected also in the present case. By considering a parabolic camber profile $\eta = \eta_0 \left( 1 - (2x / c)^2 \right)$, with $\eta$ the deviation between the chord line and the mean camber line and $\eta_0$ the maximum deviation at the middle of the wing chord, one obtains that \cite{newman1977marine},

 \begin{equation}
 \Delta \alpha = 4 \pi \eta_0 / c.
		\label{added_lift}
 \end{equation}

 We use a circular camber profile that is equivalent to Eqn. (\ref{added_lift}) as a first order approximation. In the explored parameter phase space the added angle of attack due to camber curvature is in the range of $0^{\circ}$ to $5^{\circ}$. The blade element model obtained from (2)-(3) is implemented in Python and solved numerically (see link in Supplemental Material). For each set of geometric parameters, vertical velocity $U$, and rotation rate $\Omega$, the model computes the resulting moment $M$ and vertical force $F$.

 \section{Results}

 \begin{figure}[]
 \begin{center}
  \includegraphics[width=.45\textwidth]{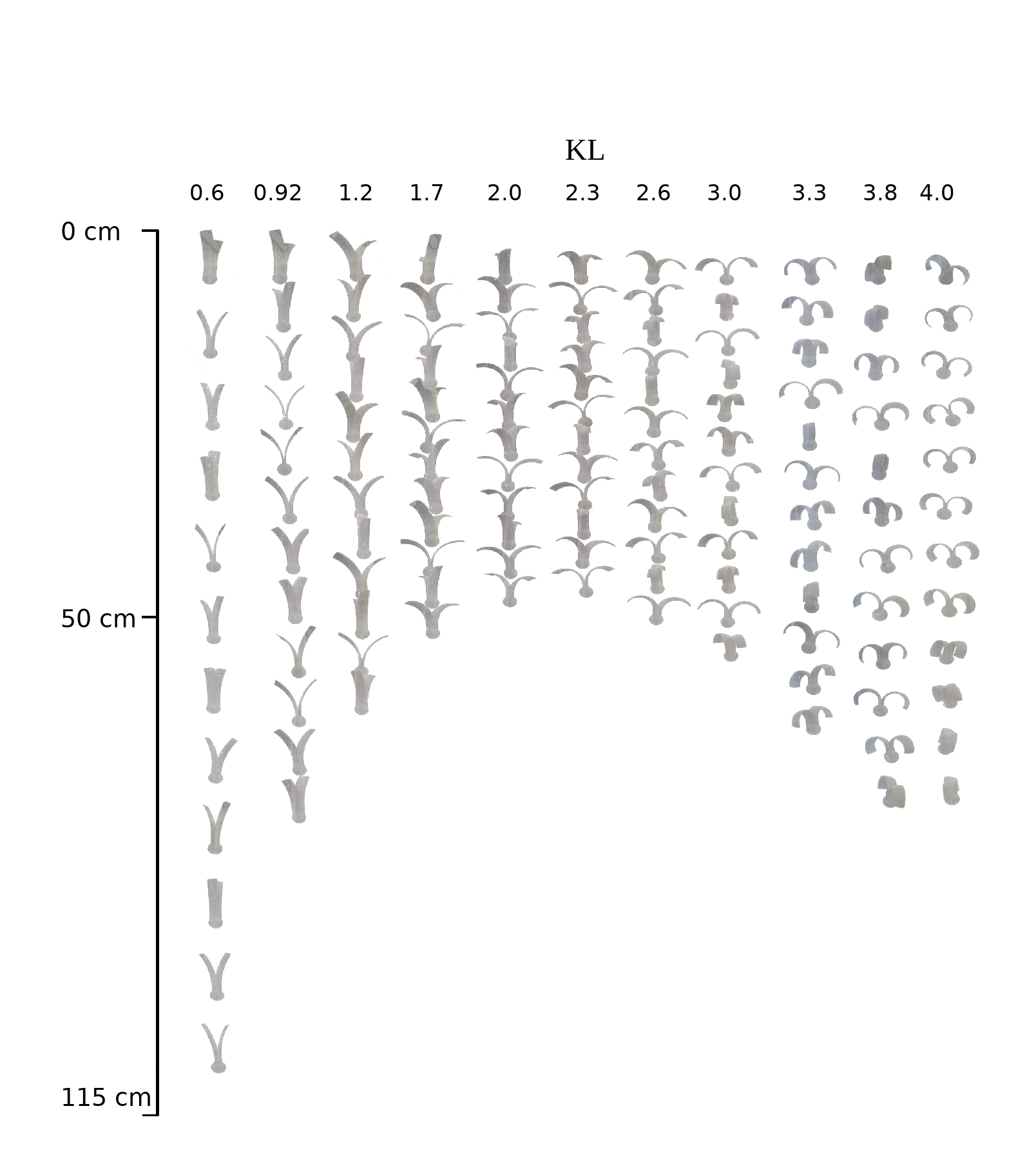}
  \caption{\label{stob0}Eleven 3D-printed synthetic double winged fruits with $\psi = 0^\circ$, $\Delta \alpha=2.5^\circ$ and the same weight $mg=20mN$, but different fold angle $KL$ are simultaneously released in the water tank. The image is a stroboscopic view of the path of the fruits and shows that the minimum terminal descent velocity is around $KL \approx 2.3$, in accordance with predictions of the blade element model (see Fig. \ref{modelcomp} (a,b)).}
 \end{center}
 \end{figure}
 
 \begin{figure*}[]
 \begin{center}
   \subfigure[ ]{\includegraphics[width=.31\textwidth]{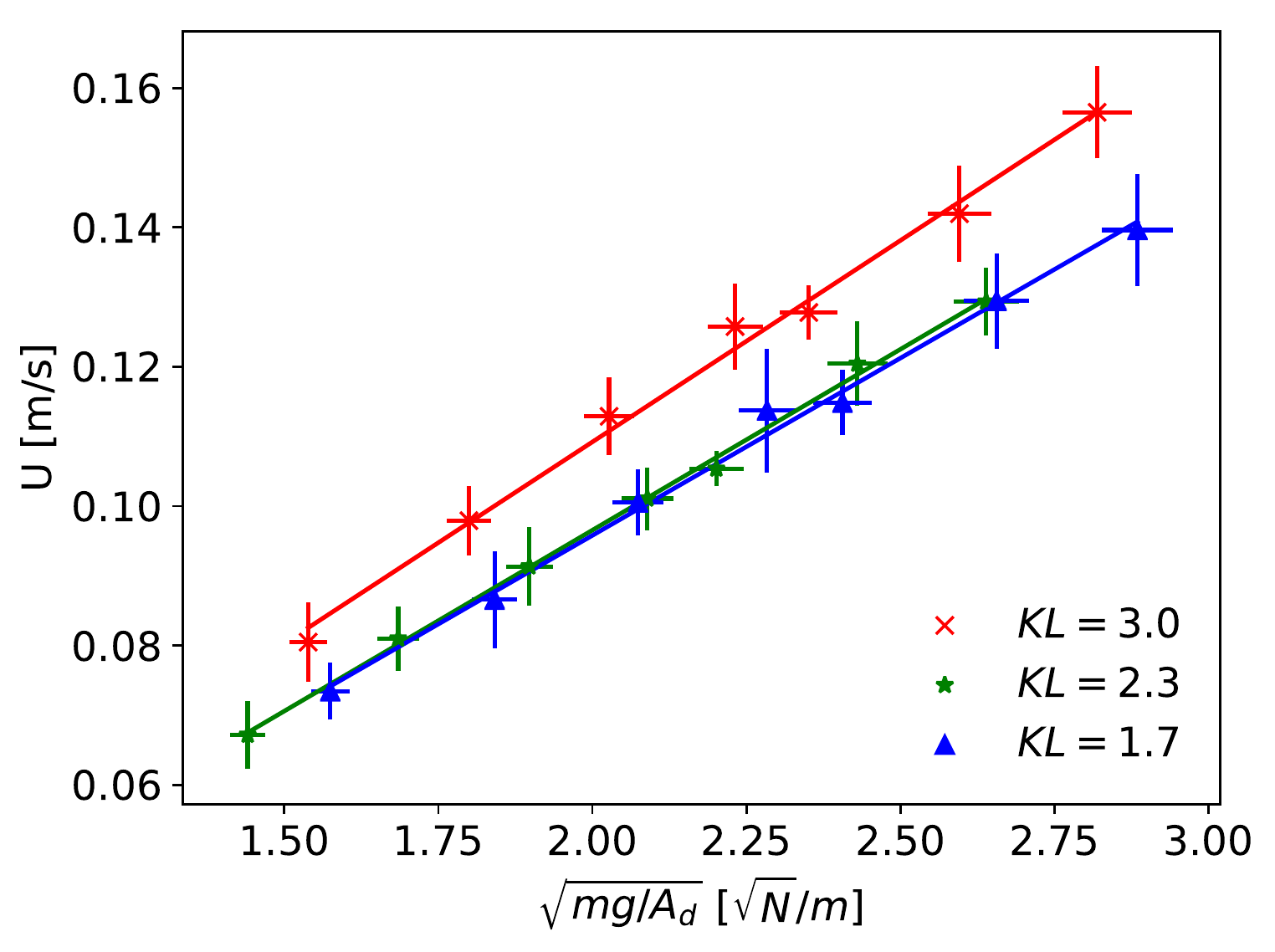}}
   \subfigure[ ]{\includegraphics[width=.31\textwidth]{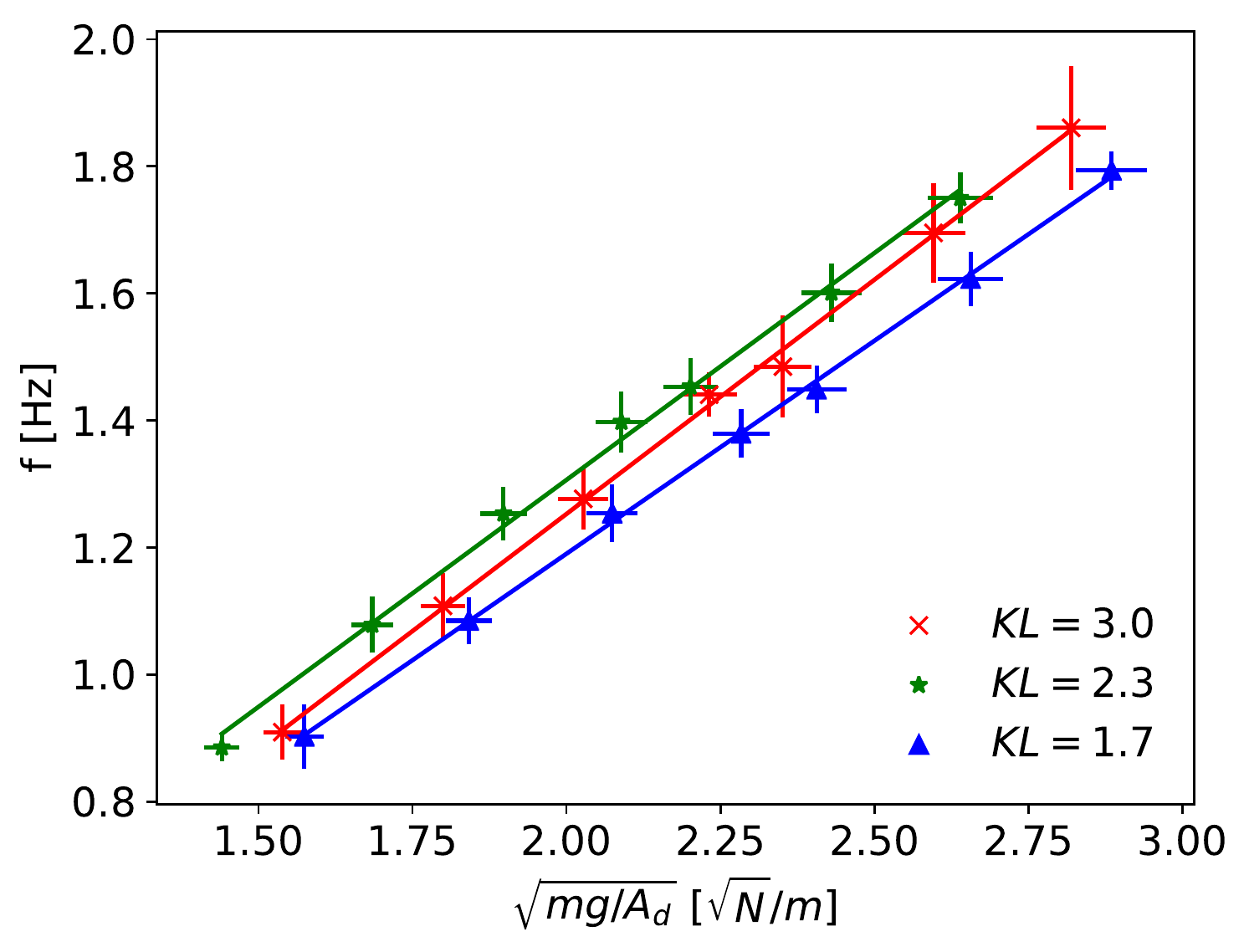}}
  \subfigure[ ]{\includegraphics[width=.31\textwidth]{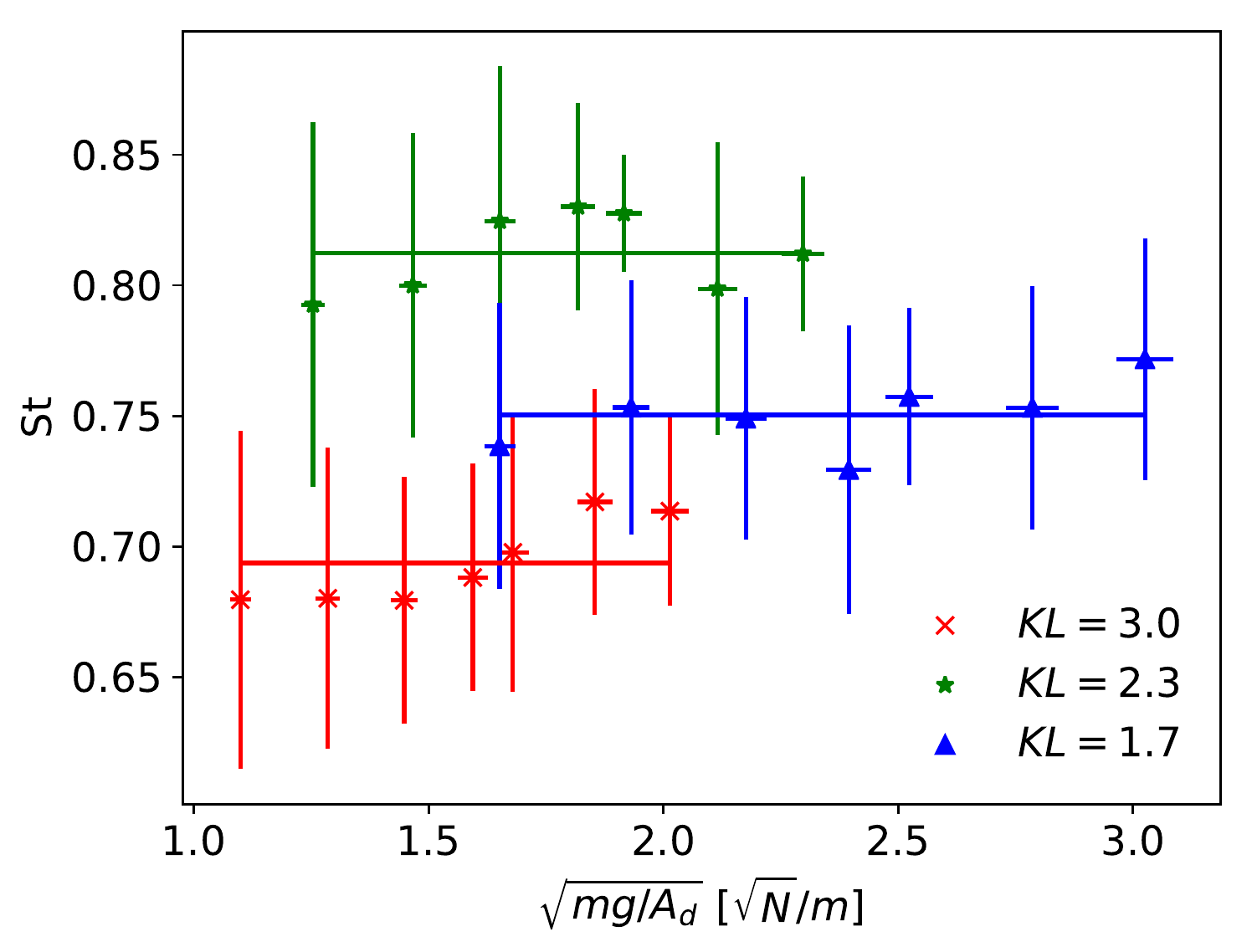}}
  \caption{\label{rescaled_fig2_3} {We plot the experimental data for $\psi=0^{\circ},\Delta \alpha=2.5^{\circ}$ for $KL = [1.7, 2.3, 3.0]$ where the wings span an area $A_d=[0.0058, 0.0069, 0.0061]m^2$, respectively. (a) The terminal descent velocity $U$ is plotted as a function of the wing loading $\sqrt{mg/A_d}$, where we see that Eqn. (\ref{eq:scaling}) does not collapse the data onto a single curve with respect to the scale curvature $KL$ where we want to note that the similarity of $KL=1.7$ and $KL=2.3$ is a special case. (b) The fruit's rotational frequency $f$ is plotted as a function of $\sqrt{mg/A_d}$, where $f$ is non-monotonic with respect to the fold angle $KL$. (c) By rescaling $U$ and $f$ we plot the Strouhal number $St$, which is constant for each geometry but the magnitude of $St$ is a function of $KL$. 
}}
 \end{center}
 \end{figure*}
\subsection{Experimental results}
We have performed an extensive experimental parameter study where we systematically alter the base wing angle $\psi$, the fold angle $KL$ and the weight of the fruit by combining 3D-printing of synthetic fruits and measurements in a water tank. The experimental phase space span;  $KL\in [0-4]$radians, $Re\in[2.4-12]\times 10^3$,  $St\approx 0.1-0.9$ and $\psi=[0^{\circ}, 35^{\circ}]$. 

\subsubsection{Base wing angle $\psi= 0^\circ$ and wing camber $\Delta \alpha = 2.5^\circ$}
We fix the base wing angle to $\psi= 0^\circ$ and the wing camber $\Delta \alpha = 2.5^\circ$, while we vary the weight $mg$ and the fold angle $KL$ in the experiments. To illustrate the flight paths of these synthetic fruits we show the stroboscopic photo in Fig. \ref{stob0}, where we have simultaneously released $11$ 3D-printed fruits at the same height and with the same weight in the water tank and track their descent distance for a fixed time $t=5.5$ s. It is clear that there is an optimal geometry to maximise the flight time. {We have compared the fold angle of these synthetic fruits with 27 species that reside in Africa, Asia, and the Americas where they were collected in the wild, which are found to have a fold angle mean(KL) = 1.8 $\pm$ 0.18, close to the optimum. This suggests that similar wing shapes have evolved in Nature and indicates that aerodynamic performance i.e., minimal descent velocity, may improve the fitness of these plants in an ecological strategy.}

In Fig. \ref{rescaled_fig2_3} (a)-(c) the measured terminal descent velocity $U$, the rotational frequency $f$ and the $St$ number are plotted as a function of the wing loading $\sqrt{mg/A_D}$ for three fold angles. It is clear from Fig. \ref{rescaled_fig2_3} (a) that the scaling law \ref{eq:scaling} does not fully capture the influence of the fold angle $KL$ as it does not collapse the data onto a single curve, but illustrates that for each individual wing geometry the terminal descent velocity scales as $\sim \sqrt{mg}$, consistent with experiments on wild fruits \cite{Augspurger1986,smith2015predicting}. The rotational frequency $f$ is also a function of $KL$ and by re-plotting $U$ and $f$ through $St=fL/U$ we see that for each geometry $St$ is constant (see Fig. \ref{rescaled_fig2_3} (c)). Thus, our experiments follow the predictions from the blade element model and our theoretical analysis (see next section , Eqn. (\ref{Eqn:csst_St})). Note that the error-bar in $St$ is $\approx 5\%$ and accumulated from the measurement error in both $U$ and $f$. As there is a clear coupling between $KL$ and $U$ we are curious to determine if there is a geometry that minimizes the terminal descent velocity, which we argue to be optimal in terms of the seeds dispersion potential. 

\subsubsection{Base wing angle $\psi= 35^\circ$ and wing camber $\Delta \alpha = 2.5^\circ$}
\begin{figure}[h!]
 \begin{center}
\includegraphics[width=.45\textwidth]{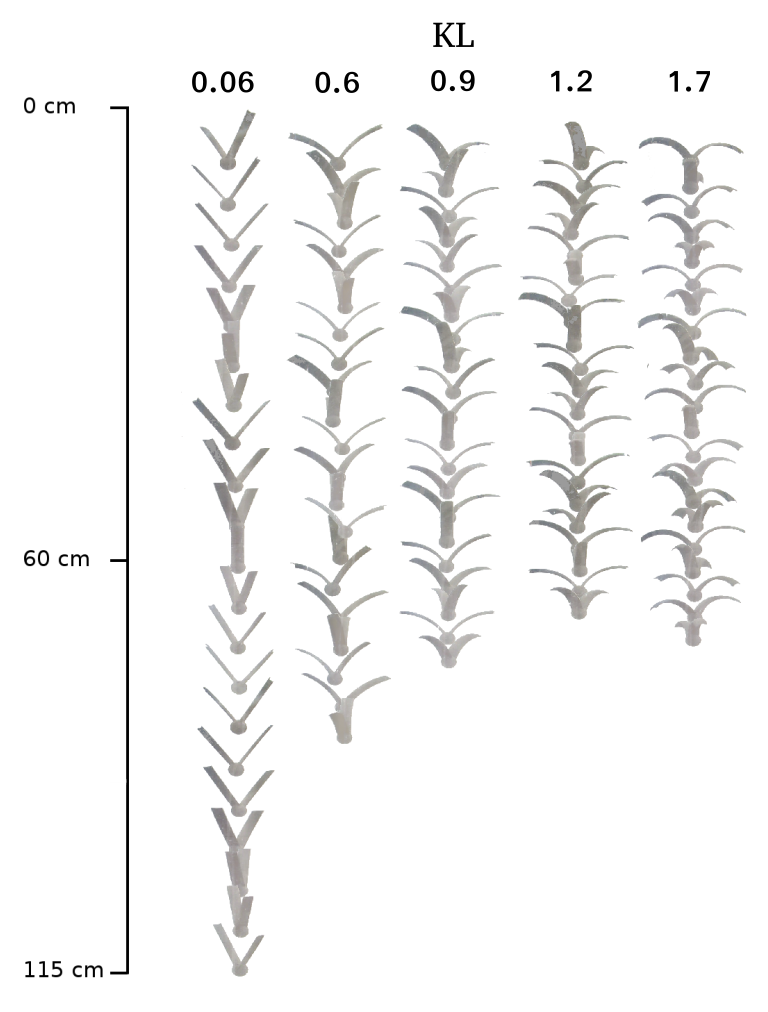}
  \caption{\label{fig:psi_35} Five 3D-printed synthetic double winged fruits with $\psi = 35^\circ$, $\Delta \alpha=2.5^\circ$ and with the same weight $mg=20mN$, but with different fold angle $KL$ are simultaneously released in the water tank. The image is a stroboscopic of the path of the fruits and shows that the minimum terminal descent velocity is around $KL = 1.2$, in accordance with predictions of the blade element model (see Fig. \ref{modelcomp}(c,d)).}
 \end{center}
 \end{figure}

Fruits and seeds can also have wings with an attachment/base angle $\psi$ that is not necessary zero. To understand how $\psi$ influences the terminal descent velocity we performed additional measurements with $\psi = 35^\circ$, $mg=19$mN and $\Delta \alpha = 2.5^\circ$ where we vary ${mg}$ and $KL$. A stroboscopic image of these flight paths are shown in Fig. \ref{fig:psi_35} for a fixed mass and flight time $t=5.5s$, to further illustrate the relationship between the $KL$ and $U$. Compared with the case when $\psi = 0^\circ$ we notice that the optimal fold angle has decreased, with a minimum terminal descent velocity obtained near $KL\approx 1.2$. 

\begin{figure*}[]
 \begin{center}
 \subfigure[ ]{\includegraphics[width=.31\textwidth]{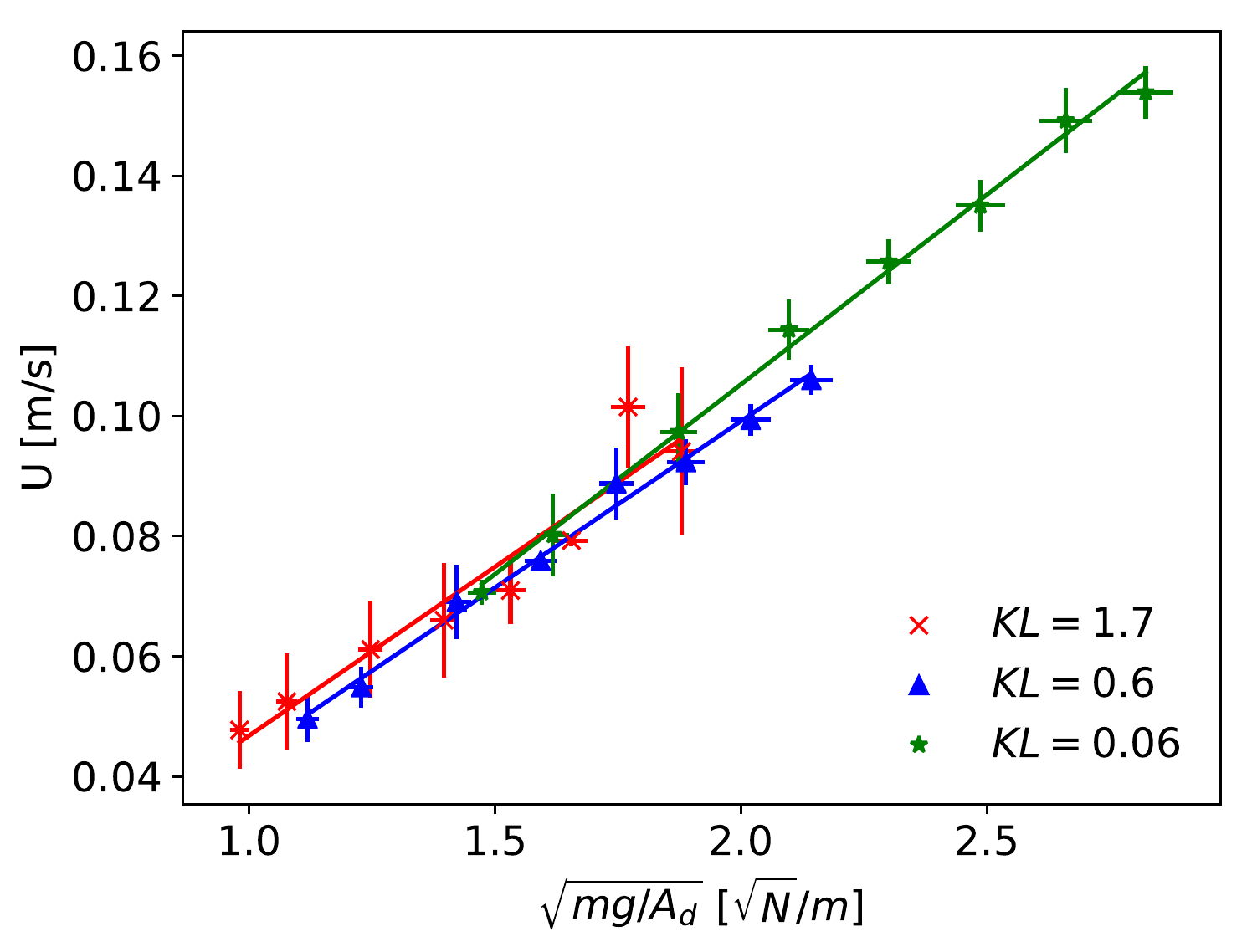}}
   \subfigure[ ]{\includegraphics[width=.31\textwidth]{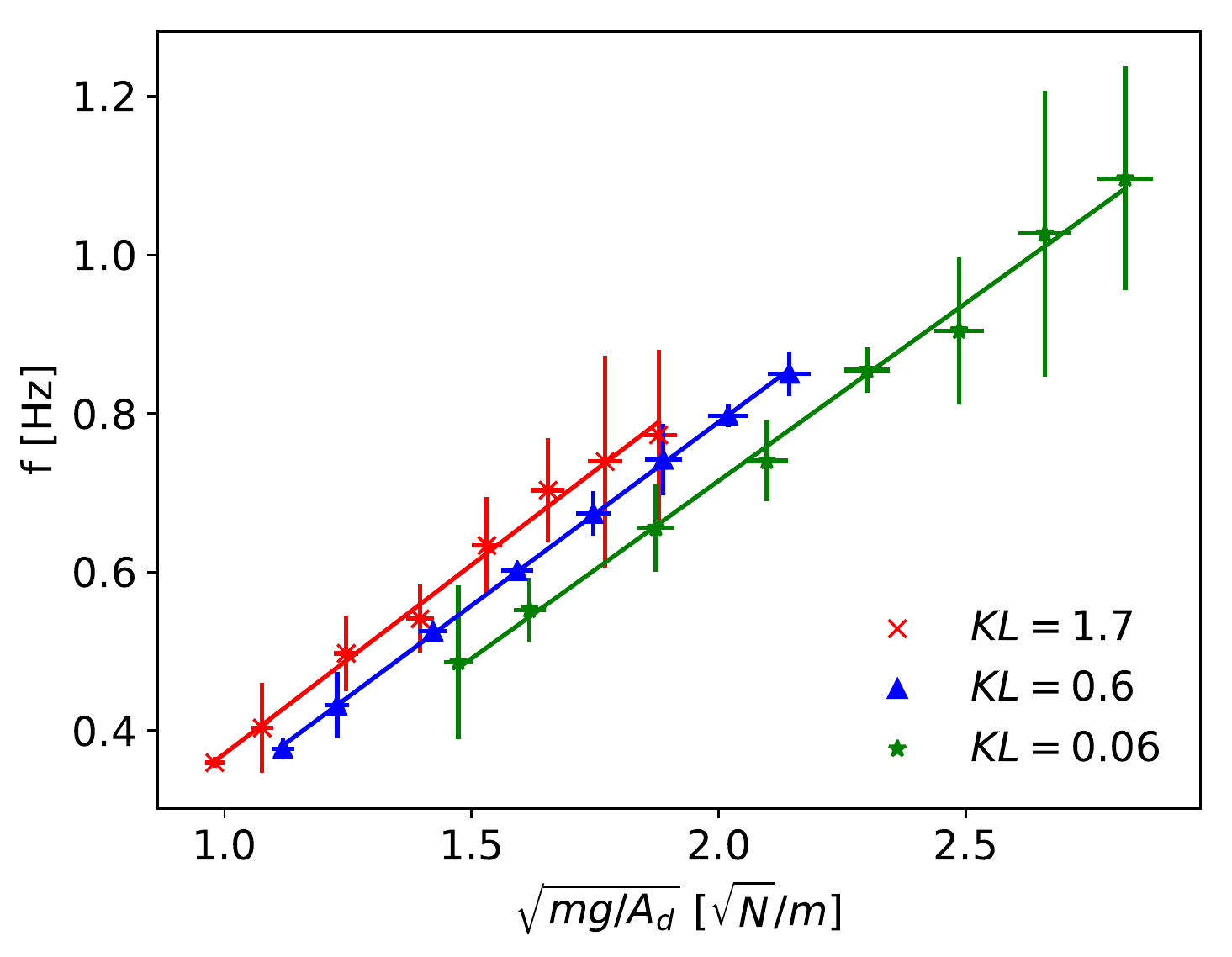}}
  \subfigure[ ]{\includegraphics[width=.31\textwidth]{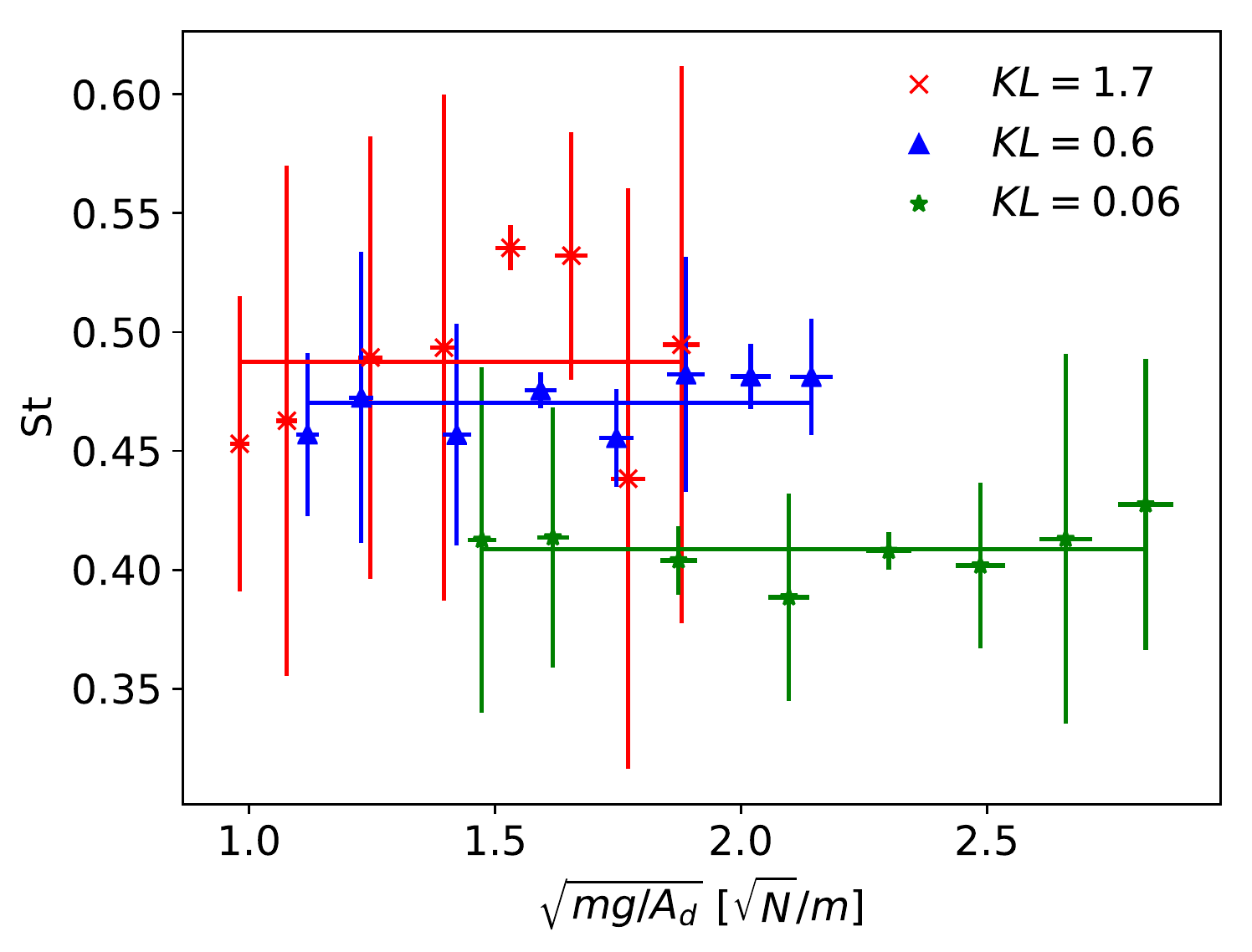}}
  \hspace{2cm}
    \caption{\label{rescaled_fig2_35} {We plot the experimental data for $\psi=35^{\circ},\Delta \alpha=2.5^{\circ}$ for $KL = [0.06, 0.6, 1.7]$ where the wings span an area $A_d= [0.0053, 0.0092, 0.012]m^2$, respectively. (a) The terminal descent velocity $U$ is plotted as a function of the wing loading $\sqrt{mg/A_d}$, where we see that Eqn. (\ref{eq:scaling}) provides a fairly good description of $U$ when the wings are not highly curved. (b) The fruit's rotational frequency $f$ is plotted as a function of $\sqrt{mg/A_d}$, where $f$ increases with the fold angle $KL$. (c) By rescaling $U$ and $f$ we plot the Strouhal number $St$, which is constant for each geometry but the magnitude of $St$ is a function of $KL$. 
    }}
 \end{center}
 \end{figure*}

Similar to $\psi=0^{\circ}$, we see that there is an influence in $U$ and $f$ as a function of $KL$ (see Fig. \ref{rescaled_fig2_35} (a)), however the influence is not as pronounced for these smaller fold angles where the wings are not that highly curved and Eqn. (\ref{eq:scaling}) predicts $U$ reasonably well. By rescaling $U$ and $f$ through $St$, we notice that it is constant for each geometry but the magnitude depends on $KL$. Experiments with additional wings, but keeping the total wing area constant were found to only make a slight change in $Re$ and $St$ numbers, and as such we did not include these here.

\subsection{Blade element model}
\begin{figure*}
 \begin{center}
  \subfigure[]
  {\includegraphics[width=.45\textwidth]{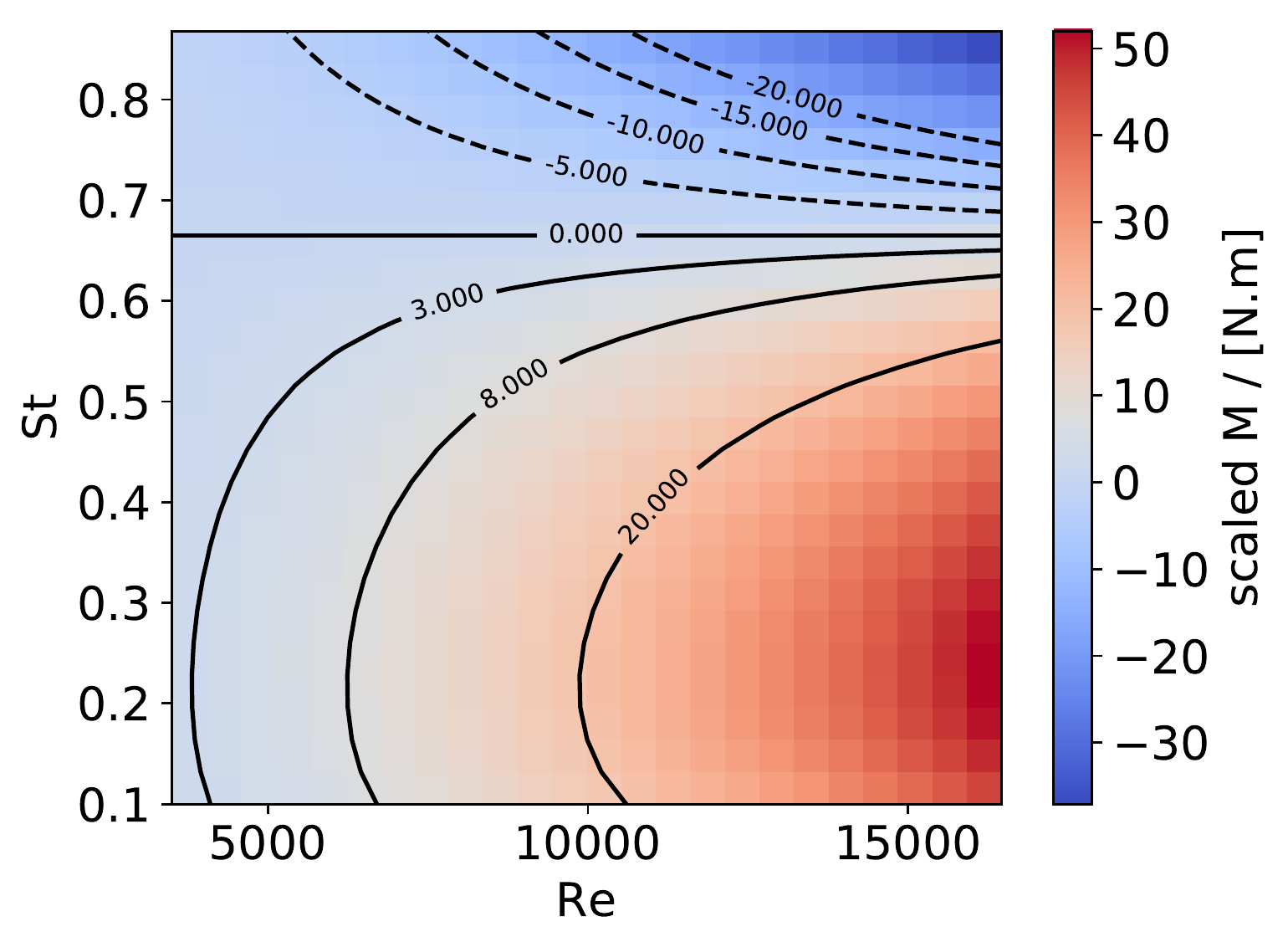}}
  \subfigure[]
  {\includegraphics[width=.45\textwidth]{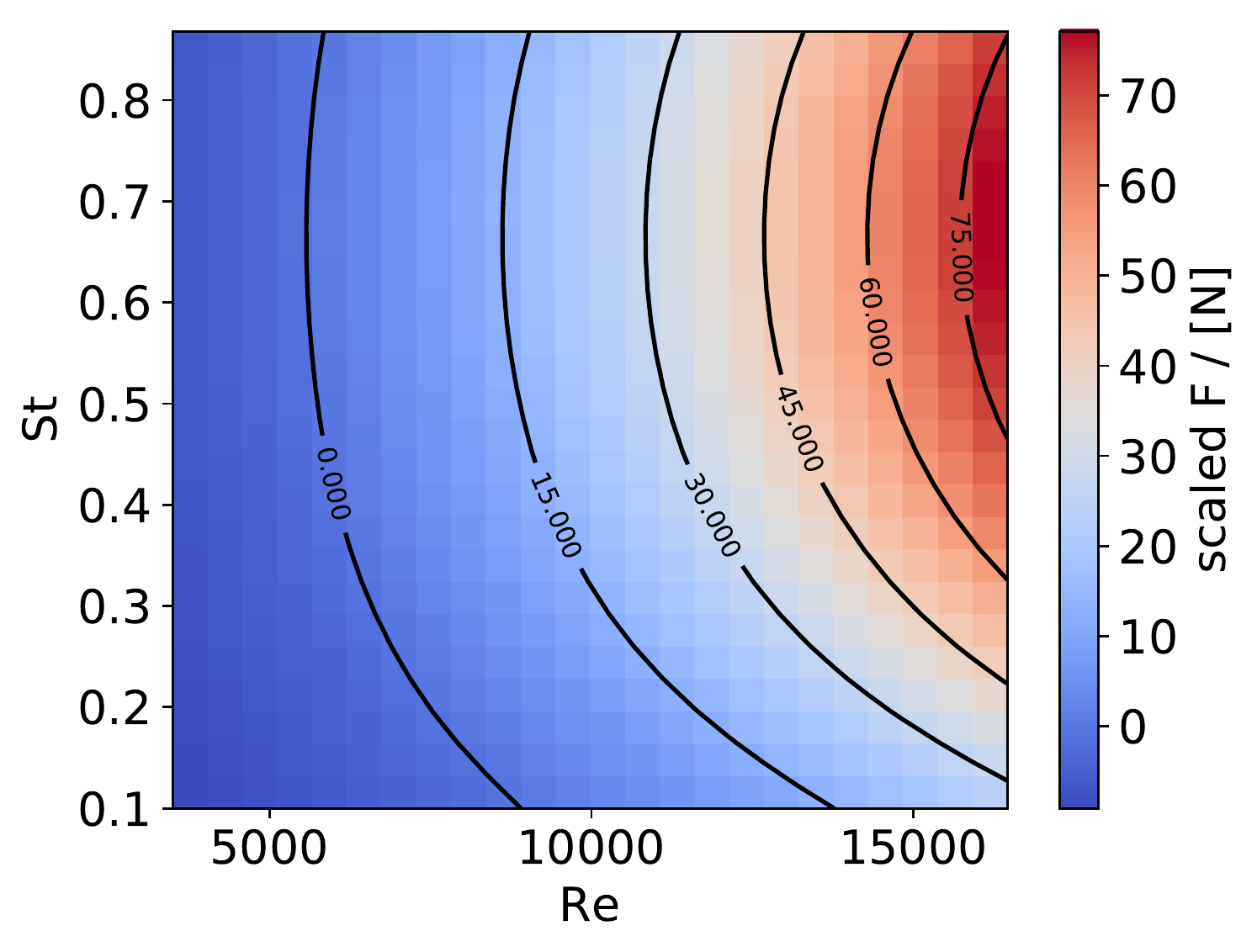}}
  \caption{\label{2D_maps} 2D {contour} maps for (a) the moment $M$, and (b) the resultant vertical force $F$ predicted by the blade element model {(\ref{Eqn:force_moment}) as a function of Re and St} for the baseline fruit geometry, with a net mass loading of 1.2 grams. The data presented in the figures are scaled of an arbitrary factor for ease of representation. The steady state descent is obtained at $F = M = 0$, and it is stable as discussed in the text.}
 \end{center}
 \end{figure*}

\begin{figure*}
 \begin{center}
   \subfigure[]
   {\includegraphics[width=.45\textwidth]{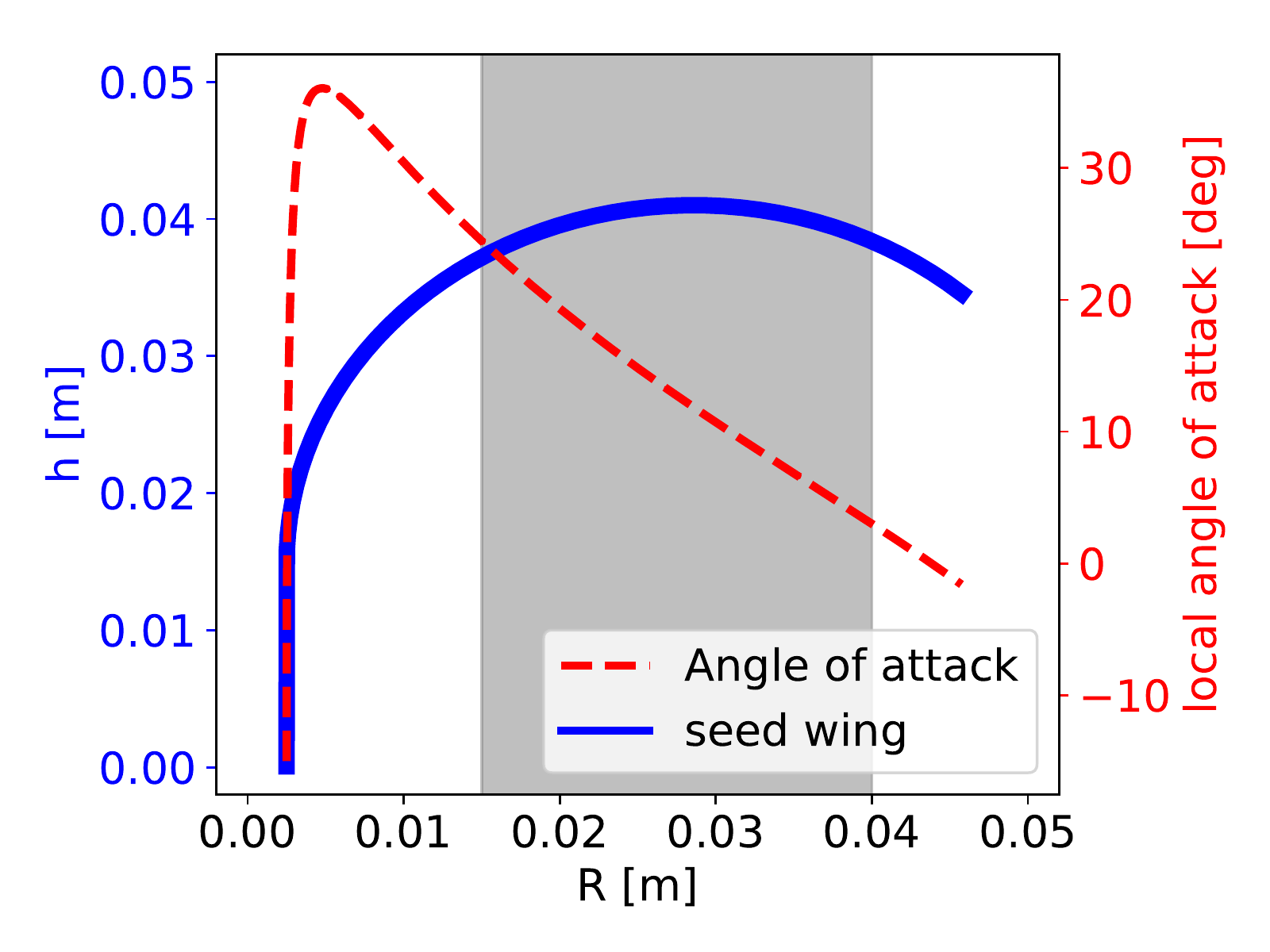}}
   \subfigure[]
   {\includegraphics[width=.47\textwidth]{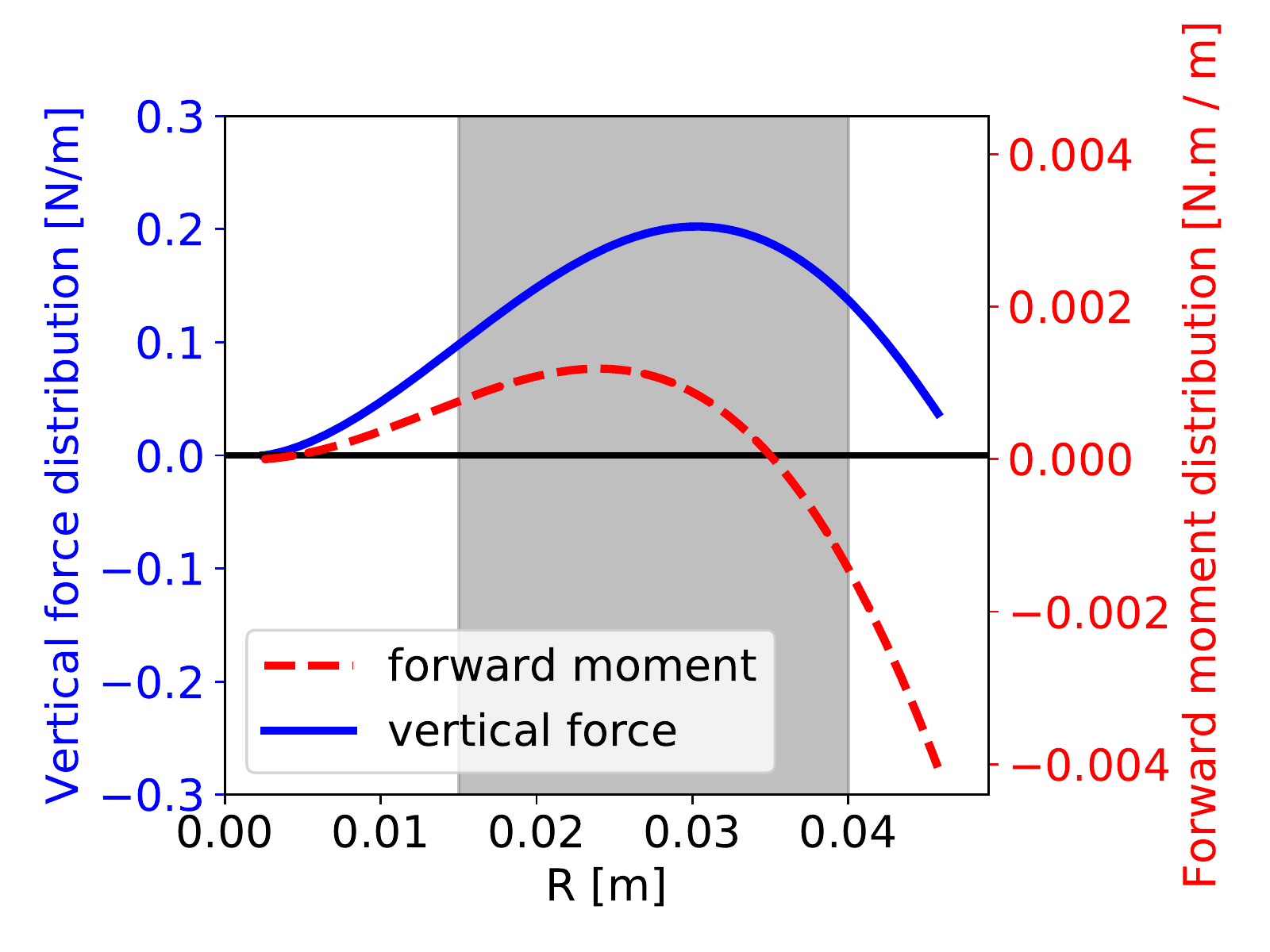}}
   \caption{\label{example_distribution} {(a) The cross section of the wing profile (left axis) corresponding to a synthetic fruit having fold angle KL = 2.3 as a function of the distance R between the blade element and the fruit's axis of rotation, and the corresponding angle of attack in the stationary regime (right axis). The part of the wing where significant lift is created is highlighted in gray, (b) the corresponding distribution of vertical force (left axis) and driving momentum (right axis).}}
 \end{center}
 \end{figure*}

 We solve the blade element model for the synthetic fruit with a model geometry with {$\psi=0^{\circ}$}$, \alpha_p=15^{\circ}$ and with a mass loading of $1.2$ grams. The data is presented in non-dimensional 2D contour maps for $F$ and $M$ as presented in Fig. \ref{2D_maps}. {{The dependence of the angle of attack, forward moment, and vertical force on the position along wingspan for a fruit with a fold angle of $KL = 2.3$ is shown in Fig. \ref{example_distribution}}}.

 In Fig. \ref{2D_maps} the level line for zero total moment is straight, corresponding to a constant Strouhal number St, only determined by the geometry of the fruit wings. This is consistent with all our experiments (Fig. 7(c), 9(c)), and can be seen in the blade element model from Eqn. (\ref{Eqn:force_moment}) by substituting the value of $F_F$:

 \begin{equation}
 M = \int_{l=0}^{l=S} \frac{1}{2} \rho U_R^2 A \left[ C_L(\alpha) \sin(\theta) - C_D(\alpha) \cos(\theta) \right] R dl.
 \label{Eqn:csst_St}
 \end{equation}

 Using a constant $St$ implies that the relative wind direction, and therefore angle of attack $\alpha$, remains constant for each blade element. Therefore, if the fruit is experiencing a zero mean torque, changing the descent speed for a given fruit geometry and $St$ will result in replacing $U_R$ by a scaled value in Eqn. (\ref{Eqn:csst_St}) while keeping all other terms constant, i.e., the resulting moment $M$ will still be zero. {This result is independent of the parametrization used for both $C_L(\alpha)$ and $C_D(\alpha)$, and verifies that the blade element model captures well the main features of the flow physics.}

\begin{figure*}[ht!]
 \begin{center}
  \subfigure[]
  {\includegraphics[width=.45\textwidth]{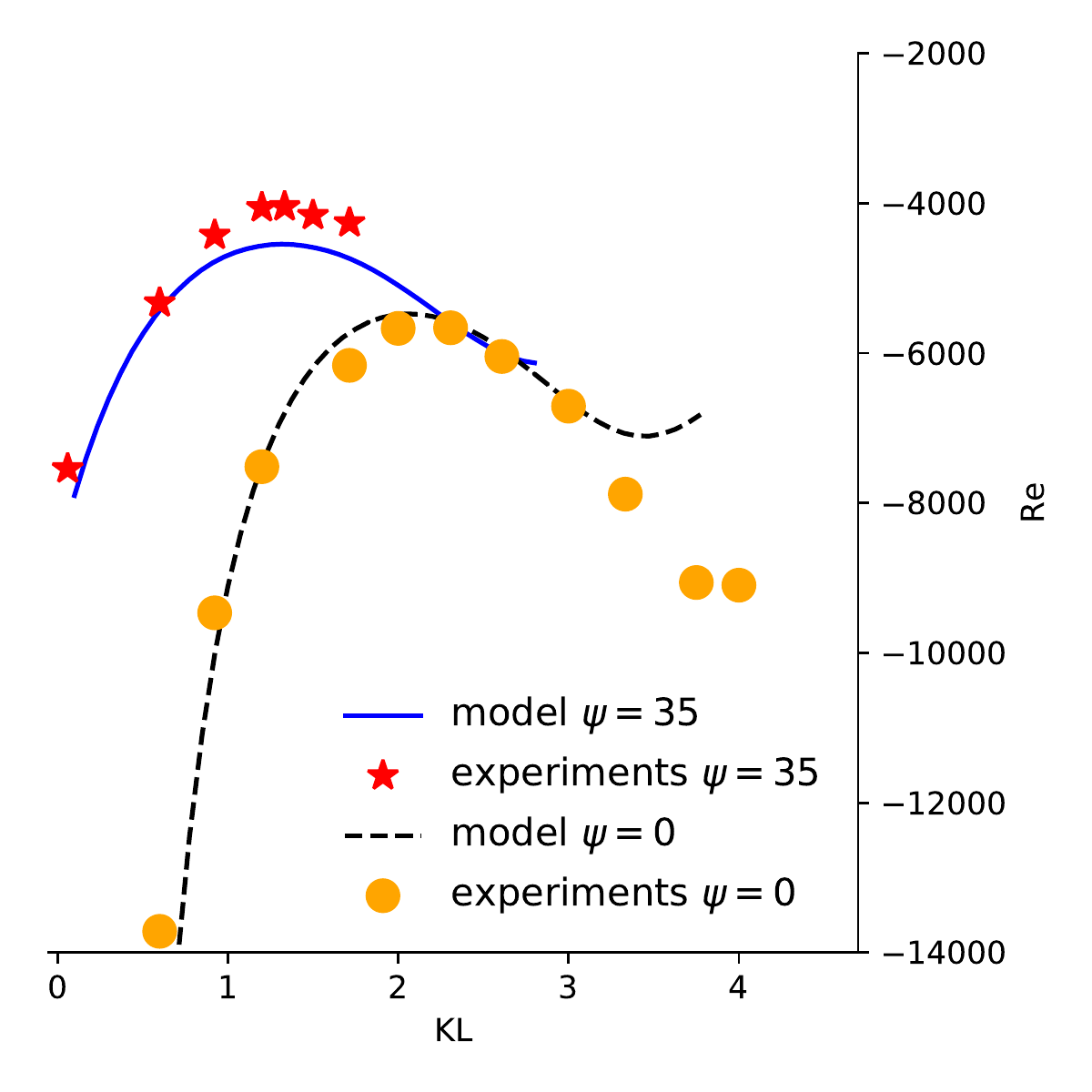}}
  \subfigure[]
  {\includegraphics[width=.45\textwidth]{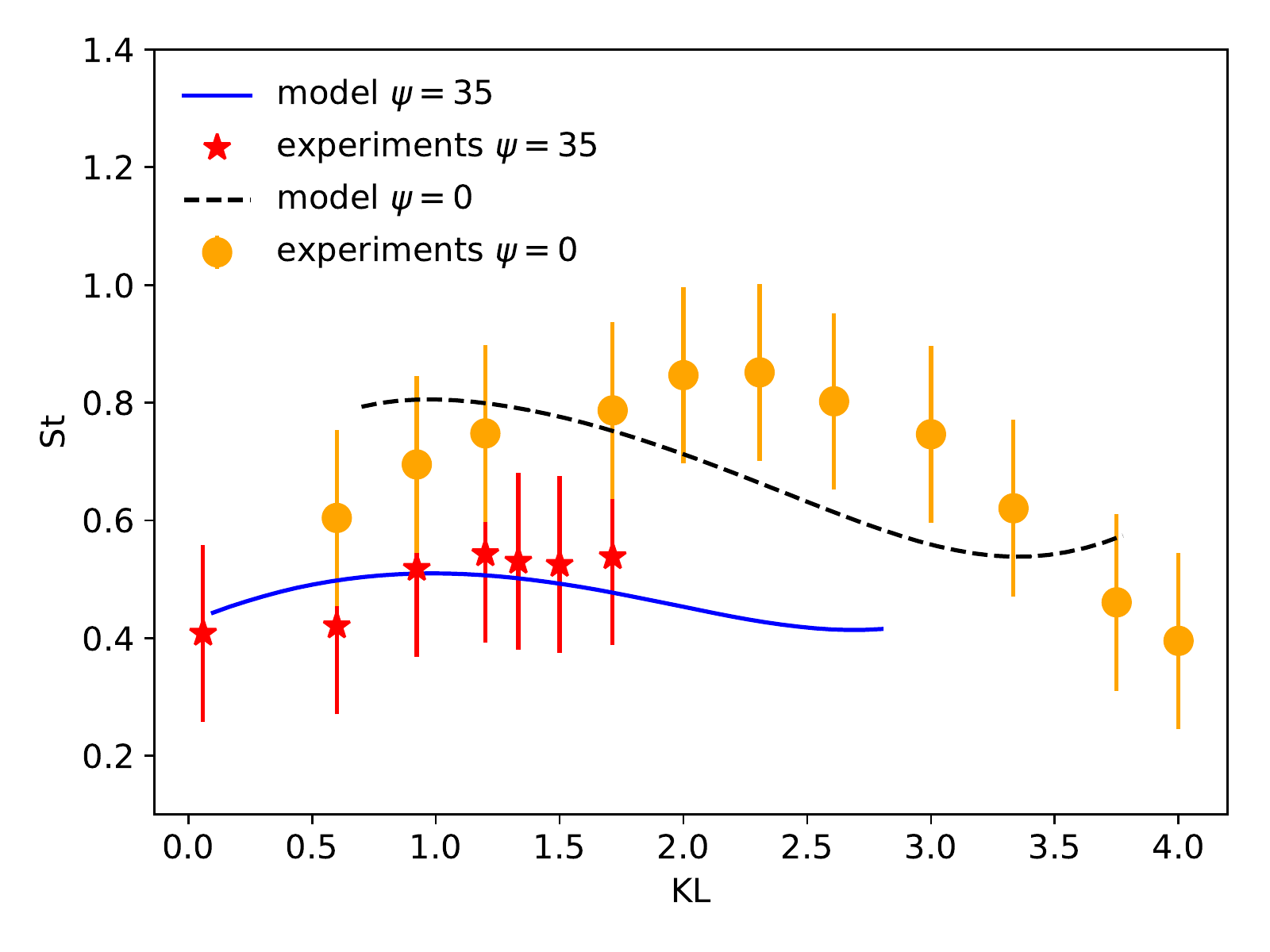}}

  \caption{\label{modelcomp} {(a) We solve the blade element model using the parameterization shown in Fig. \ref{CL_o_CD}, for the two series of seeds already presented in Fig. 6 and 8. The experimental data is shown by the markers, which illustrates that the fold-angle $KL$ is the essential parameter to obtain a minimal terminal descent velocity that would suggest optimal wind dispersion potential, as previously discussed. Good agreement is found regarding the general shape of the curves, the position of the optimal KL for each base angle $\psi$, and the relative change of falling velocity observed between Fig. 6 and 8. (b) We show the Strouhal number St, for the same set of seeds. Here, satisfactory agreement is found when considering the simplicity of the model. Both the typical magnitude of St in each case, the effect of modifying the base angle $\psi$, and the general trend of St with respect to KL are satisfactorily reproduced by the model.}}
 \end{center}
 \end{figure*}

 In order to find the terminal descent velocity and rotation rate of the falling fruit, we need to determine the parameters [$U$, $\Omega$], in non-dimensional form, [$Re$, $St$] for which the total resultant vertical force $F$ and the resulting moment $M$ are zero. This is done numerically, based on an analysis of the 2D maps for both quantities. The equilibrium point is stable, as confirmed by the experimental measurements. Indeed, as visible in Fig. \ref{2D_maps}, for a large $St$ we get a negative moment, i.e., a reduction of $St$, while a small $St$ leads to a positive moment, i.e., an increase of $St$, implying that the fruit returns to equilibrium. Similarly, the vertical force increases with a large descent velocity $U$ i.e., a large $Re$ number, leading to an upwards resulting force that slows down the vertical motion of the fruit, and opposite for a small descent velocity leading to a large downwards resulting force, which will increase the descent speed. Direct comparison between the experimental results and the predictions from the blade element model show that they are in good agreement, given the phenomenological nature of this model (Fig. \ref{modelcomp}).
 
\begin{figure*}[ht!]
 \begin{center}
  \subfigure[]
  {\includegraphics[width=.45\textwidth]{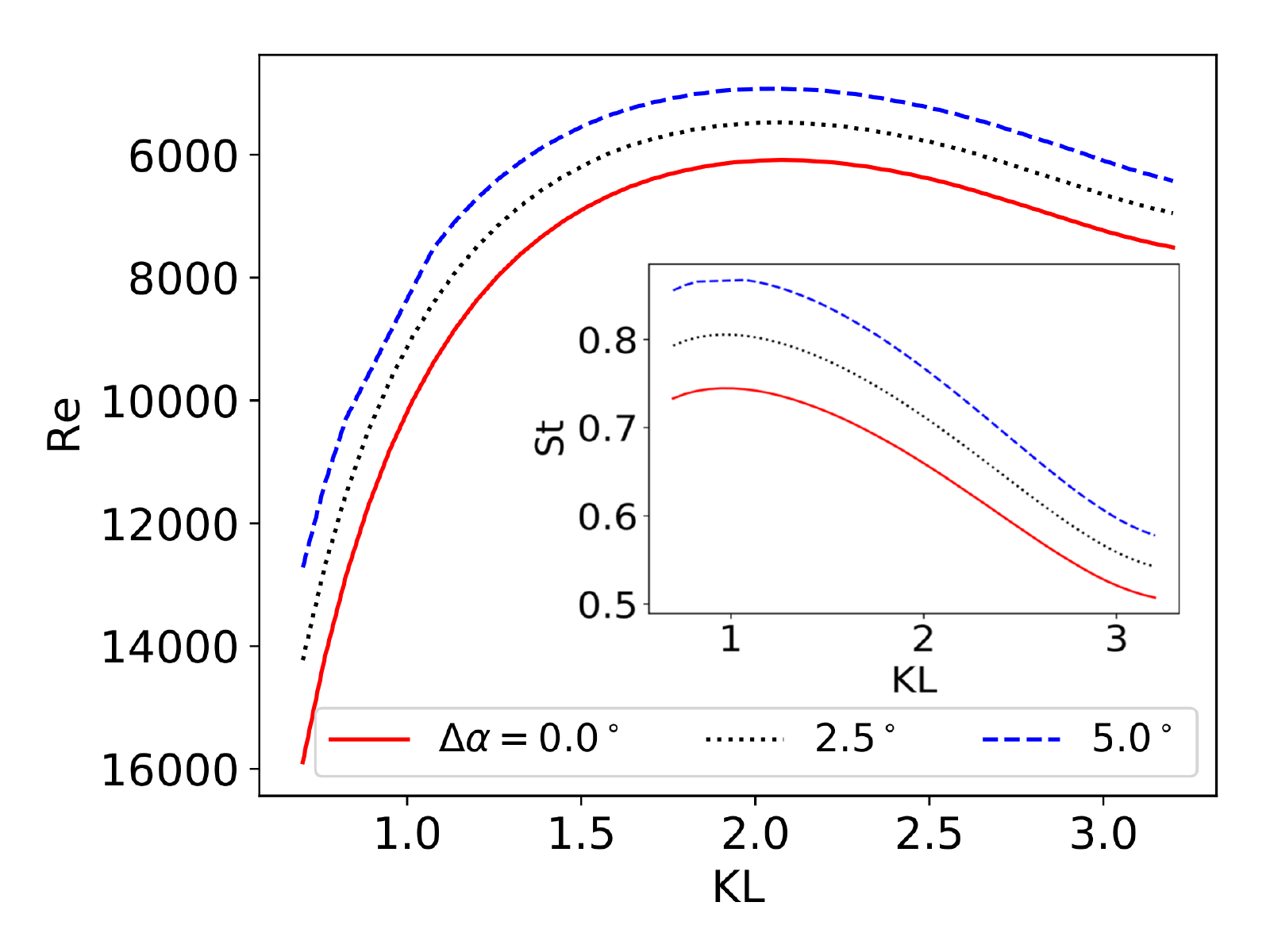}}
  \subfigure[]
  {\includegraphics[width=.45\textwidth]{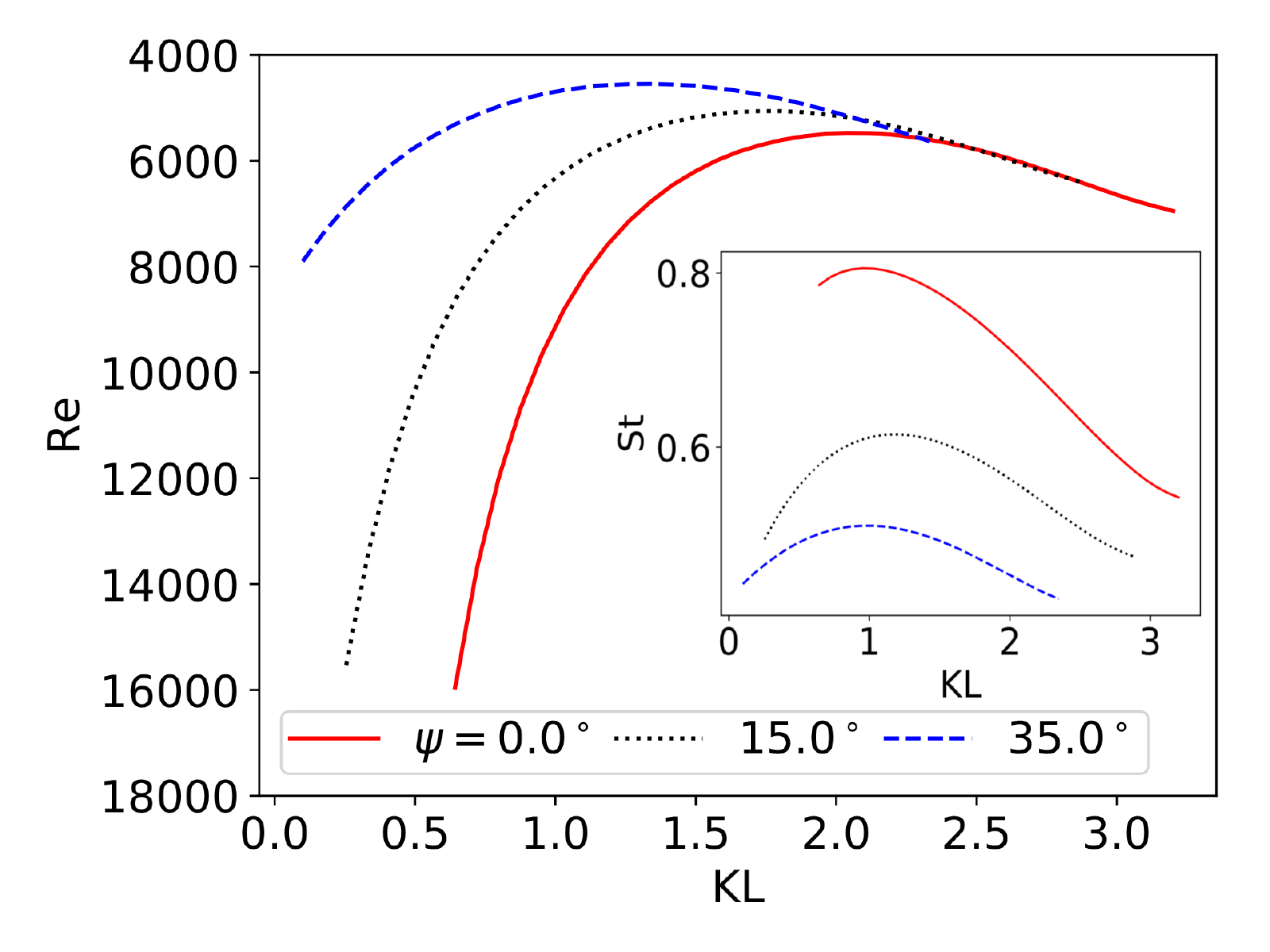}}
  \caption{\label{effect_other_params} (a) The terminal descent velocity and Strouhal number as a function of sepal camber ($\Delta \alpha$) and sepal fold angle, as predicted by the blade element model for $\Psi=0^{\circ}$. The addition of a camber does not affect the placement of the minimal descent velocity along the $KL$ axis, but only influences its magnitude. This is due to improved autogiration, as visible through the inset for the Strouhal number. (b) Changes in the base angle $\psi$ formed between the base of the wing and the vertical axis shift the placement of the minimal descent velocity on the $KL$ axis, which is moved to smaller $KL$ when $\psi$ increases. Simultaneoulsy, a larger $\psi$ leads to a larger projected wingspan which increases the relative wind created due to rotation on the outer part of the wings, which in turn reduces the equilibrium Strouhal number.}
 \end{center}
 \end{figure*}

To demonstrate the sensitivity in geometrical changes in the wing, we show in Fig. \ref{effect_other_params} the effect of both $\Delta \alpha$ and $\psi$ on the minimal terminal descent speed as predicted by the blade element model. The camber $\Delta \alpha$ has no influence on the placement of the minimal terminal descent velocity along the $KL$ axis, but only influences the magnitude of the descent velocity as it increases/decreases the effective lift force. The base angle $\psi$ in Fig. \ref{effect_other_params}b has both a limited influence on the minimal terminal descent velocity, and shifts the position along the $KL$ axis, consistent with our experiments see Fig. 6, 8: a larger $\psi$ leads to smaller value fold angle $KL$ to ensure a minimal descent velocity.

{Indeed, for a fixed weight we see that as we start with a KL value close to 0 and gradually increase the fold angle the descent velocity $U$ is being reduced, see Fig. \ref{modelcomp}. For $\psi=0$ a minimal $U$ is identified for $KL\approx 2.0$ and as we further increase $KL>2.0$ the descent velocity starts to increase. We can in part understand the placement of these optimal fold angles if we infer that the total force in the vertical direction will scale with the projected length of the wing on the horizontal axis as seen from Eqn. (1). However, this cannot alone explain the shift in minimal terminal descent speed as shown in Fig. \ref{rescaled_fig2_3} (a). As $KL$ increases so does the wing swept area and the wing tip approaches an approximately horizontal shape. Near the peak in the $KL-Re$ the wing tip is close to beeing horizontal, and as $KL$ increases the horizontal part on the wing moves towards the wing base/fruit, which then reduces the relative velocity as $R$ is smaller along with the vertical lift force, although the wing swept area is nearly the same. The wing shape must also influence how vortices are shedded, which generate circulation and lift. Therefore, these curved wings must be sufficiently long to have horizontal segments, but also sufficiently short to ensure that their tip segments are primarily aligned along the horizontal direction. However, the exact optimum for the fold angle KL is a result of the complex interplay between the flow and the wing geometry.}
 
\section{Conclusions}

We have presented an experimental study {and a minimal flow model encompassing the main physical ingredients} of the flight of synthetic whirling fruits, which mimic those found in nature. Our results point to  geometrical shapes of the wings of multi-winged seeds, fruits and diaspores, which provide them with an optimal dispersion potential, i.e., maximal flight time, and compares favourably with wing geometries found in the wild \cite{PRL}. For whirling fruits to maximize the time they are airborne, their appendages that function as wings must not curve too much or too little. Our methodology consists of a combination of rapid prototyping by 3D-printing, a minimal theory and experiments, and may be adopted for other studies of how wing geometry affects flight, which may help understand the evolutionary links between form and fitness of flight organs found in Nature. 

 \section{Acknowledgements}
We are very thankful to James Smith, Jaboury Ghazoul, Yasmine Meroz, Renaud Bastien and Anneleen Kool for stimulating discussions about {\em Dipterocarpus} fruits and for their valuable input on this work. We thank Olav Gundersen for the assistance with developing the experiment design. We gratefully acknowledge financial support from the Faculty of Mathematics and Natural Science, and the UiO:LifeScience initiative at the University of Oslo.

\section{Supplemental Material}
\subsection{Source files; CAD files, images of wild fruits, and the numerical code solving the blade element model}

Additional material including the baseline CAD file used for performing the parametric analysis, the images of wild fruits with osculating circle, and the code to solve the blade element model, is available at the following address: \url{https://github.com/jerabaul29/EffectFoldAngleAutorotatingSeeds}.

\subsection{Results from the blade element model with $C_D(\alpha)$ and $C_L(\alpha)$ inspired by flate plate measurements}

 \begin{figure*}[h]
 \begin{center}
  \subfigure[]
  {\includegraphics[width=.45\textwidth]{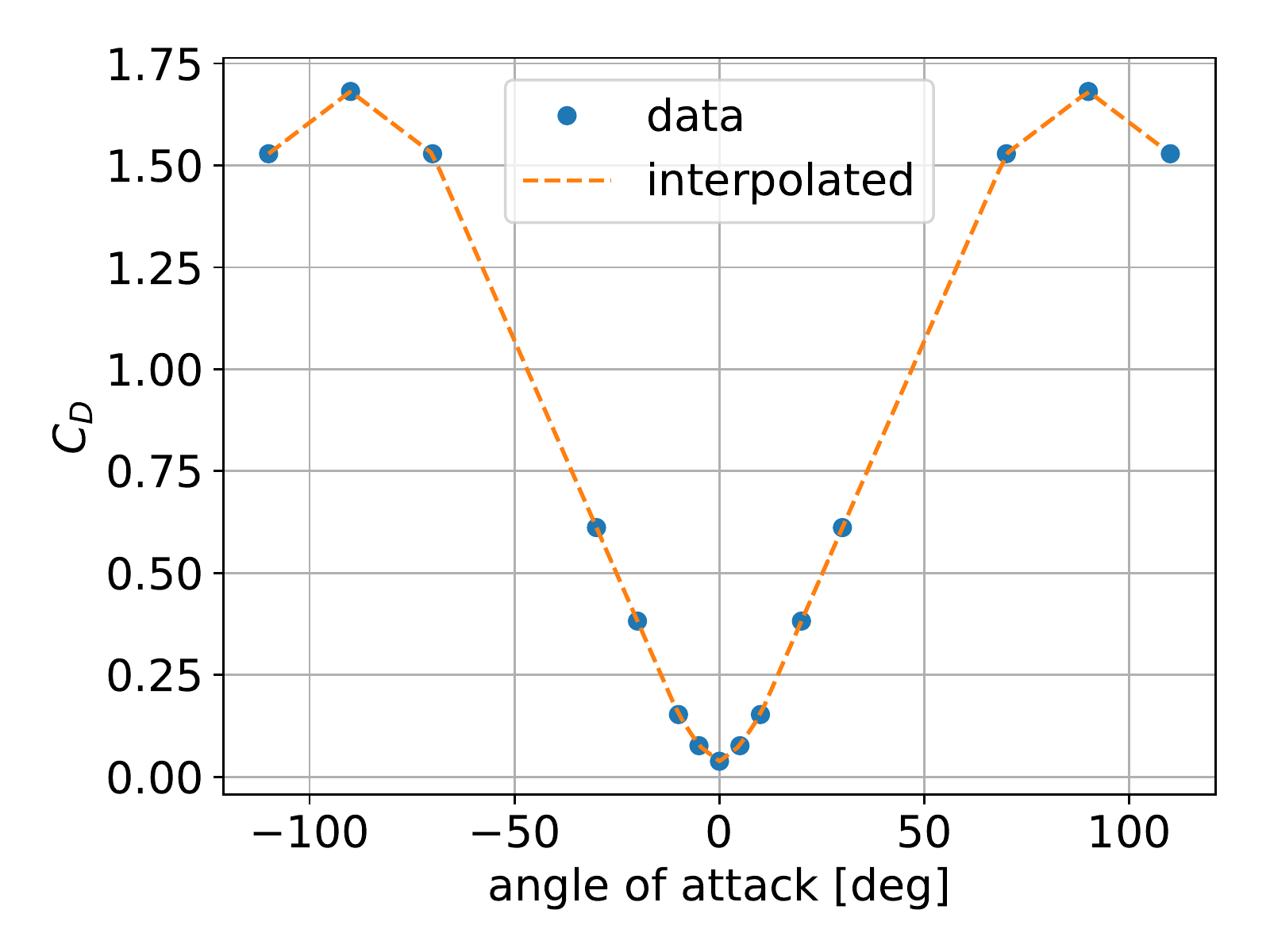}}
  \subfigure[]
  {\includegraphics[width=.45\textwidth]{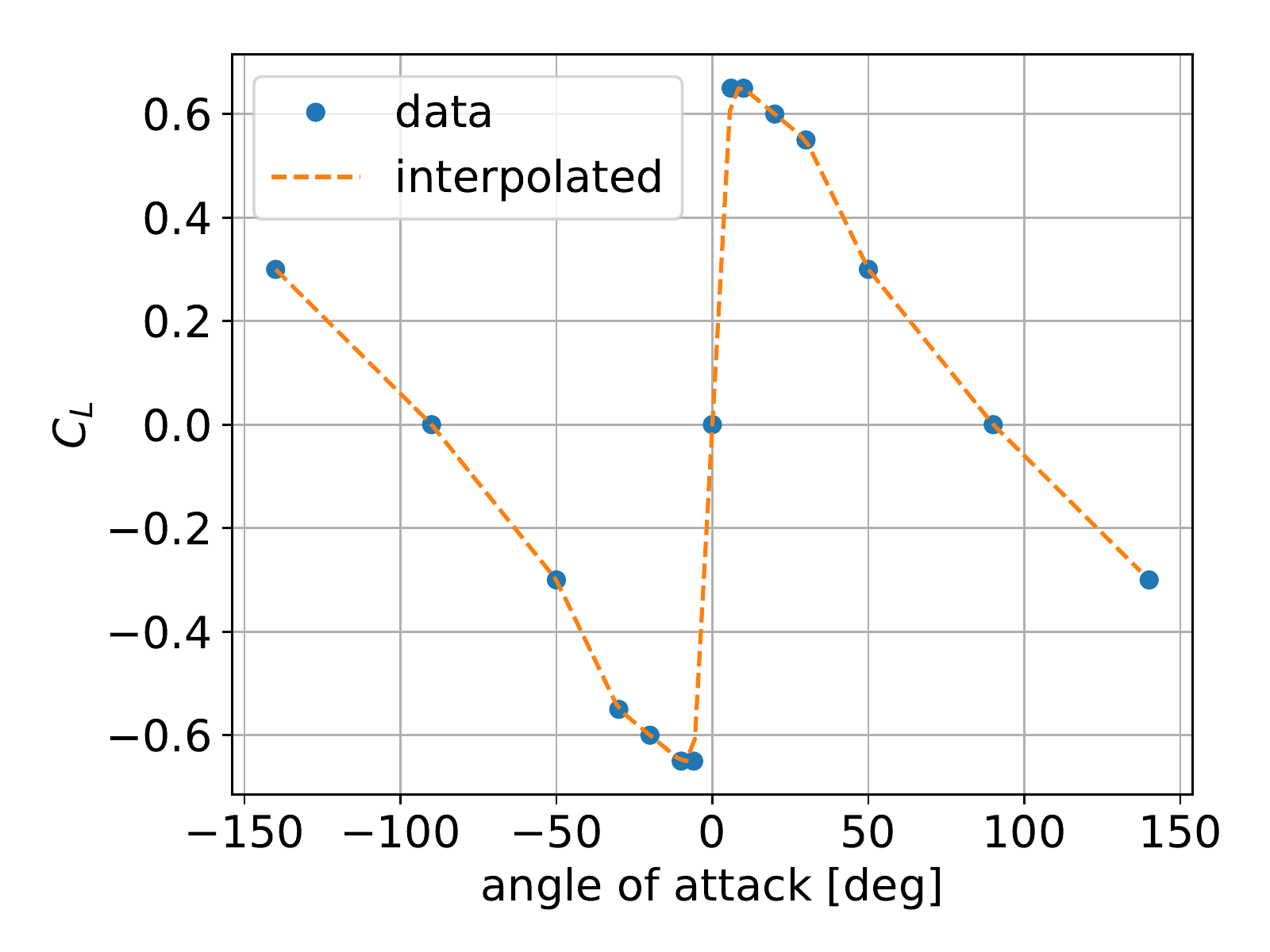}}
  \caption{\label{old_cd_cl} Parametrization for (a) $C_D(\alpha)$, and (b) $C_L(\alpha)$ inspired from translating wings at slightly higher Re numbers. A sharp stall behaviour is present for an angle of attack of around $6^{\circ}$.}
 \end{center}
 \end{figure*}

  \begin{figure*}[h]
 \begin{center}
  \subfigure[]
  {\includegraphics[width=.45\textwidth]{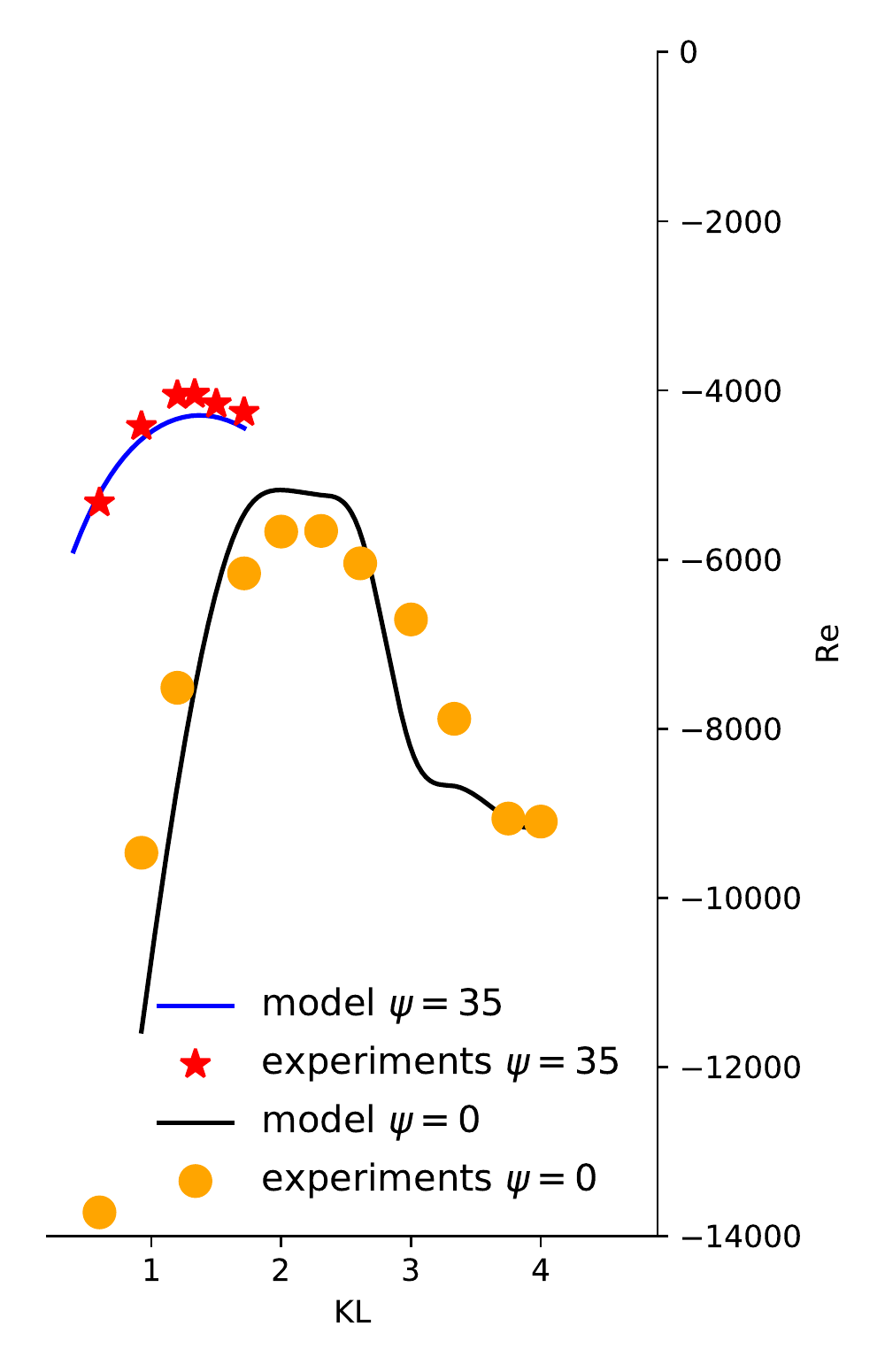}}
  \subfigure[]
  {\includegraphics[width=.45\textwidth]{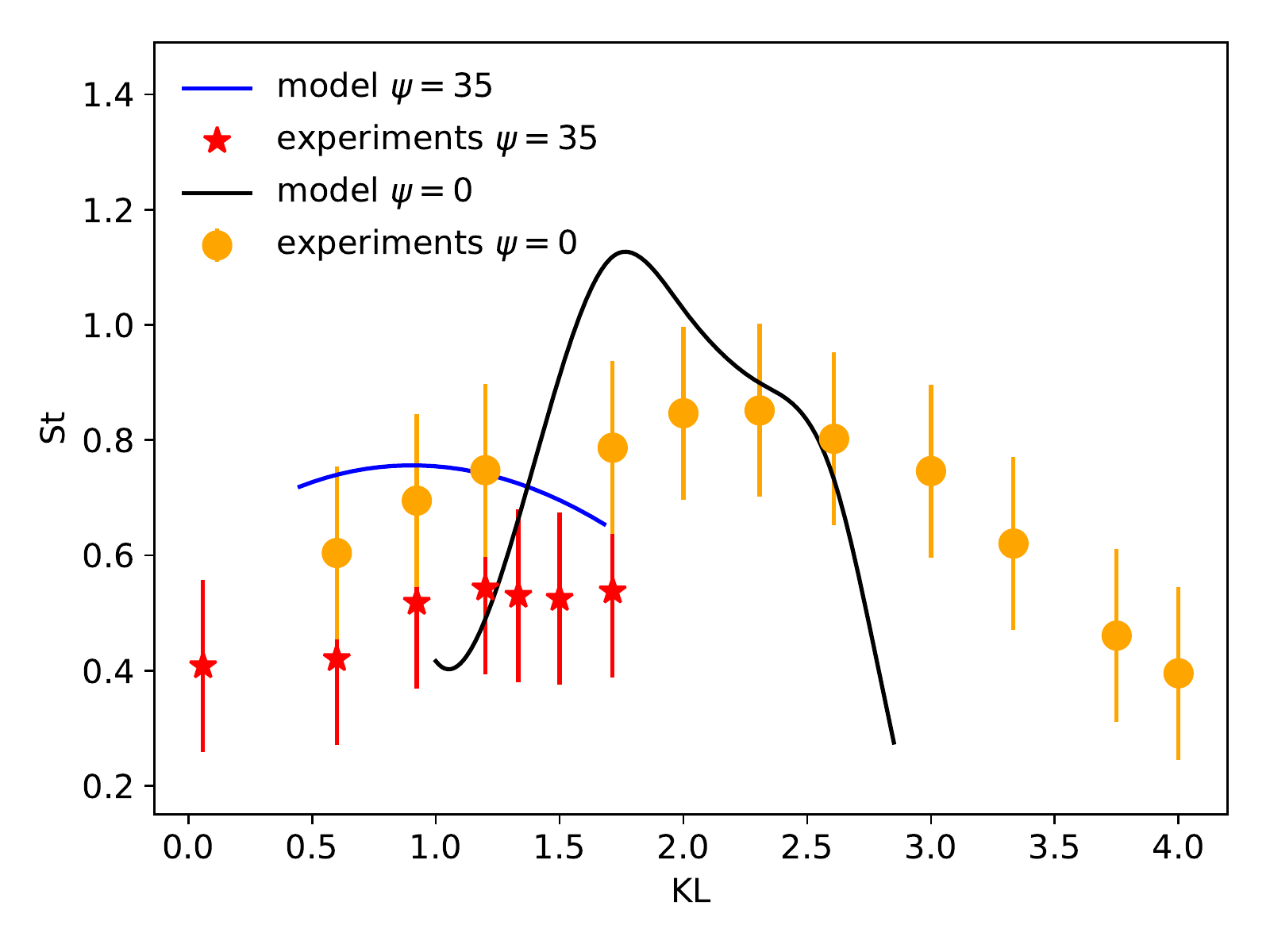}}
  \caption{\label{comp_old_parametrization} Comparison between experimental results with synthetic 3D printed bio-mimetic fruits and the blade elements model, using the parametrization of Fig. \ref{old_cd_cl}. The results are in reasonable agreement for the terminal velocity (Reynolds number, panel (a)). The results for the Strouhal number (panel (b)) follow the same trend as the experimental results, though the agreement is not as good as in Fig. \ref{modelcomp}, especially at higher values of $KL$ when stall takes place.}
 \end{center}
 \end{figure*}
 
 \begin{figure*}[h]
 \begin{center}
  {\includegraphics[width=.45\textwidth]{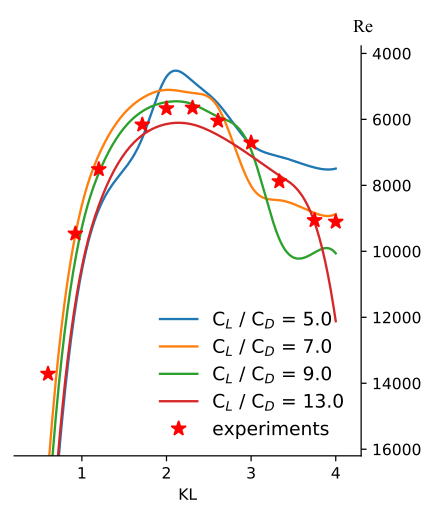}}
  \caption{\label{effect_cl_cd_legacy} Influence of varying the peak value of $C_L(\alpha) / C_D(\alpha)$ for the parametrization inspired by translating wings. While the details of the curves shape is altered, the position of the optimum curvatures remains mostly unchanged.}
 \end{center}
 \end{figure*}

In this Supplemental Material, we present the results obtained by using a parametrization with stall inspired from translating wings at slightly higher Reynolds (Re) numbers as described in the main section of the text \cite{Wang449, Lentink2705}, corresponding to the $C_L$ and $C_D$ coefficients from Fig. \ref{old_cd_cl}. Results are obtained by solving the blade elements model, in the same way as in the article. As visible in Fig. \ref{comp_old_parametrization}, the curves for the Re number are in good agreement with the experiments, however overall the results obtained with the parametrization inspired from rotating wings at smaller Re numbers. The Strouhal (St) number is also mostly following the trends observed in experiments, though the discrepancy with the experiments is larger than what was reported in Fig. \ref{modelcomp}, especially for $\psi = 0$ when the strong stall takes place (high values of KL). Finally, the results for the terminal Re number are found to be insensitive to the peak value of $C_L(\alpha) / C_D(\alpha)$, as shown in Fig. \ref{effect_cl_cd_legacy}.

 \end{document}